%% file: Software-2026-reactive-hyper.tex
\documentclass[AMA,Times1COL]{WileyNJDv5} 

\DeclareTextFontCommand{\texttt}{\ttfamily\fontsize{11}{13}\selectfont}
\DeclareOldFontCommand{\tt}{\normalfont\ttfamily\fontsize{11}{13}\selectfont}{\mathtt}
\makeatletter
\def\abstractfont{\hsize\abs@colii@hsize\rmfamily\fontsize{11}{15}\selectfont\leftskip7\p@\rightskip\leftskip}
\makeatother

\usepackage{listings}
\usepackage{multicol}
\usepackage{graphicx}
\usepackage{float}
\usepackage[most]{tcolorbox}
\usepackage{caption}
\usepackage{amsmath}
\usepackage{amssymb}
\usepackage{booktabs}
\usepackage{array}

\usepackage{sintax-highlight}

\articletype{Research Article}%

\received{}
\revised{}
\accepted{}
\journal{Software: Practice and Experience}
\volume{00}
\copyyear{2026}
\startpage{1}

\begin{document}

\title{BRHC: Backend-driven Reactive Hypermedia Controls with a Statically Typed Kotlin DSL}

\author[1]{Fernando Miguel Carvalho}

\author[1]{Paulo Carvalho}

\author[1]{Leonel Correia}

\author[1]{Ricardo Gomes}

\author[2]{Juho Veps\"al\"ainen}

\authormark{CARVALHO \textsc{et al.}}
\titlemark{BRHC: BACKEND-DRIVEN REACTIVE HYPERMEDIA CONTROLS}

\address[1]{\orgname{Instituto Superior de Engenharia de Lisboa, Instituto Politécnico de Lisboa}, \orgaddress{\city{Lisbon}, \country{Portugal}}}

\address[2]{\orgdiv{Department of Computer Science, School of Science}, \orgname{Aalto University}, \orgaddress{\city{Espoo}, \country{Finland}}}

\corres{Corresponding author Fernando Miguel Carvalho, \email{miguel.gamboa@isel.pt}}



\abstract[Abstract]{
    AI-assisted coding tools (e.g., Copilot, Cursor, Claude) are increasingly
    ubiquitous and enable rapid generation of web applications. However, this
    raises concerns regarding complexity, longevity and the long-term
    maintainability of generated systems. A key source of complexity is the
    heterogeneity between backend and frontend programming models, where
    multiple languages and paradigms are combined within a single application,
    often leading to duplicated logic and fragmented state management. To
    address this issue, recent approaches (e.g., HTMX, Turbo Hotwire, Datastar,
    etc.) follow the Hypermedia-Driven Application (HDA) model, positioning HTML
    as the primary communication medium between client and server. Unlike
    SPA-centric architectures, HDA systems shift the application state and
    interaction logic to the server, where backend-driven reactive signals
    synchronize with the client user interface. However, these approaches still
    introduce complexity through custom attributes and do not fully eliminate
    JavaScript, particularly in computed expressions. In this work, we propose a
    statically typed approach using a Kotlin-based HTML DSL (Domain-Specific
    Language) for backend-driven reactive web applications. We extend the
    HtmlFlow Kotlin DSL with typed custom HTML attributes (i.e., Datastar data-*
    attributes) and signal-based bindings using statically typed builders. We
    demonstrate the approach through a catalog of reactive interaction patterns
    and a Petclinic Spring MVC case study. The results indicate that the
    proposed approach can nearly eliminate the need for JavaScript while
    improving type safety and preserving a homogeneous programming model across
    frontend and backend, bridged through a backend-driven reactive,
    signal-centric architecture.
}


\jnlcitation{\cname{%
\author{Carvalho F.M.},
\author{Carvalho P.},
\author{Correia L.},
\author{Gomes R.}, and
\author{Veps\"al\"ainen J.}}.
\ctitle{BRHC: Backend-driven Reactive Hypermedia Controls with a Statically Typed Kotlin DSL.} \cjournal{\it MTI.} \cvol{2026;00(00):1--00}.}

\maketitle

\input{sections/sec01-intro}
\input{sections/sec02-related-work}
\input{sections/sec03-signal-base-dsl.tex}
\input{sections/sec04-patterns}

\input{sections/sec04-formal-def}
\input{sections/sec05-case-study}
\input{sections/sec07-performance.tex}
\input{sections/sec08-discussion.tex}
\input{sections/sec09-conclusions}

\bibliography{bibliography}

\end{document}

%% file: sections/sec01-intro.tex
\section{Introduction}
\label{sec:introduction}

The Web is the most widely accessible application platform, serving as the primary
computing environment for billions of users worldwide. According to the
International Telecommunication Union (ITU), approximately 6 billion people were
using the Internet in 2025, representing nearly three-quarters of the global
population \cite{ITUFactsFigures2025}. Furthermore, web technologies continue to
dominate modern digital services, with JavaScript being used by 98.9\% of all
surveyed websites according to W3Techs \cite{W3TechsJavaScript}. At the same time,
modern web applications have become increasingly complex, relying heavily on
client-side execution and large JavaScript codebases. The HTTP Archive Web Almanac
2025~\cite{HTTPArchive2025} reports that 98.1\% of web pages make at least one
JavaScript request, while the median mobile page weight has continued to increase
steadily over time, reflecting a substantial increase in resource requirements
over the past decade. The rapid adoption of JavaScript progressively transformed
the Web from a document navigation system into a programmable application
platform~\cite{taivalsaari2017web}. Client-side libraries such as
jQuery~\cite{jquery} simplified DOM manipulation and asynchronous communication
with servers, encouraging richer client-side interaction patterns. Techniques such
as partial page updates and
transclusion~\cite{nelson1995heart,krottmaier2001,maurer2006}—where fragments of
remote documents are dynamically inserted into existing pages—enabled more
interactive user interfaces without full page reloads. Over time, the complexity
of client-side logic grew significantly, culminating in the widespread adoption of
the Single-Page Application (SPA) model~\cite{mesbah2008}.

In recent years, there has been a renewed interest in the Hypermedia-Driven Application
(HDA)\footnote{\url{https://htmx.org/essays/hypermedia-driven-applications/}}
approach~\cite{gross2023hypermedia} as an architectural alternative to JavaScript-heavy
SPAs~\cite{vepsalainen2026revisiting}. This shift is motivated by concerns related to
complexity, steep learning curves, and long-term maintenance costs. Rather than
continuing the progressive migration of application logic to the client, these
approaches re-emphasize server-centric interaction models and aim to reduce the
reliance on client-side JavaScript where possible.

The hypermedia-driven approach builds upon the original Web architecture
principles of Hypermedia as the Engine of Application State (HATEOAS), a core
constraint of REST \cite{fielding2000architectural}. In this context, \textit{hypermedia
controls} such as links, forms, and buttons, act as the primary interaction
mechanism between client and server.
\textit{Hypermedia controls} were originally defined by Fielding as elements based
on affordances that provide users with actionable interaction choices within a
hypermedia system. Building on this foundation, Gross~\cite{gross2024hypermedia}
refined the concept into a more formal model in which a \textit{hypermedia
control} is understood as an element that (1) responds to an event trigger, (2)
issues an HTTP request to a target URL, and (3) integrates the resulting response
into the user interface at a designated location. JavaScript libraries such as
htmx~\cite{htmx}, Turbo~\cite{turbo}, and
Datastar~\cite{datastar} leverage the hypermedia control model by
extending HTML with custom attributes such as \texttt{hx-}, \texttt{turbo-}, respectively, 
or data attributes i.e. \texttt{data-} for Datastar, which augment standard HTML elements with
declarative behavior for client-server interaction. While this formalization
clarifies the interaction model of \textit{hypermedia controls}, practical implementations
are typically realized through untyped or weakly typed mechanisms embedded in
HTML.

Our work extends the HtmlFlow Kotlin DSL (Domain-Specific
Languages)~\cite{htmlflow} by leveraging Kotlin~\cite{kotlinlang} as a
statically typed host language to suppress the issues associated with untyped or
string-based expressions in hypermedia control attributes.
The proposed model eliminates JavaScript expressions
by replacing them with statically typed expressions defined directly in the
backend host language, namely Kotlin. We demonstrate the applicability of the
approach through a catalog of \textit{Backend-driven Reactive Hypermedia
Control} (BRHC) patterns (as popularized by htmx examples and also used in
Datastar) and a case study based on the well-known Petclinic Spring MVC web
application\footnote{\url{https://spring-petclinic.github.io}}.
We explore a formalization of these patterns that enables their systematic
implementation across different statically typed backend environments.
This work addresses the following research questions:

\begin{enumerate}
\item \label{rq:elimination}\textbf{RQ1:} To what extent can \textit{hypermedia
controls} and reactive signal bindings be expressed through a statically typed
Kotlin DSL while eliminating JavaScript expressions?

\item \label{rq:typesafety}\textbf{RQ2:} How does the proposed approach contribute
to type safety and the unification of frontend and backend programming models in a
Spring Petclinic case study?
\end{enumerate}

The remainder of this paper is organized as follows. Section 2 reviews related
work in modern web architectures highlighting the trade-offs between client-side
complexity and backend-driven interaction models. Section 3 introduces HTML
templating and safe reactive signal management, discussing traditional server-side
templating engines and their limitations, and motivating the use of internal DSLs
in statically typed languages to address the type-safety concerns of
RQ~\ref{rq:typesafety}. Section 4 presents Backend-driven Reactive Hypermedia
Control (BRHC) patterns, outlining the core principles for composing
backend-driven interactions using signals and \textit{hypermedia controls}, and
addresses RQ~\ref{rq:elimination} by showing how these interactions can be
expressed without application-specific JavaScript expressions. Section 5 provides
a formalization of these BRHC control patterns, defining their structural and
behavioral semantics. Section 6 evaluates the proposed approach through a
PetClinic case study, demonstrating its applicability and addressing
RQ~\ref{rq:typesafety}.
Section 7 presents a performance evaluation of three PetClinic implementations
(HtmlFlow Datastar, Thymeleaf and React). Finally, Section 8 discusses the
findings, answers the research questions, examines threats to validity and
limitations, and concludes the paper with directions for future work.

%% file: sections/sec02-related-work.tex
\section{Related Work}
\label{sec:background}

Modern web development has increasingly shifted toward client-side architectures
based on Single-Page Applications (SPAs) \cite{sireteanu2021front}. While these
architectures enable rich interactive experiences, they also introduce challenges
related to frontend complexity, duplicated business logic, and long-term
maintainability \cite{mikkonen2008web}, as well as hydration costs inherent to
client-side rendering \cite{sireteanu2021front}.

One prominent response to these limitations is the emergence of \emph{disappearing
frameworks} \cite{vepsalainen2023desapearingframeworks}. These frameworks
represent a paradigm shift in modern web development by attempting to minimize
their own presence in the final application. Rather than extending the traditional
SPA model, these approaches focus on reducing the amount of JavaScript shipped to
the client through techniques such as server-side rendering, progressive
enhancement, and selective hydration. By prioritizing minimal client-side
execution and deferring computation whenever possible, they aim to improve
performance, simplify application delivery, and reduce frontend complexity.

Related approaches have explored alternative rendering strategies to address the
costs of client-side hydration. Islands Architecture restricts hydration to
isolated interactive regions while preserving server-rendered content, thereby
reducing JavaScript execution and improving page performance
\cite{HallieOsmani2022,Miller2020}. Similarly, Incremental Static Regeneration
(ISR)~\cite{nextjs_isr_guide}, popularized by Next.js~\cite{nextjs} in 2020,
combines static site generation with selective server-side revalidation, enabling
applications to retain many of the performance benefits of static content while
supporting dynamic updates.

Another architectural response to SPA complexity is the adoption of
micro-frontends \cite{geers2020micro}, which decompose large frontend
applications into independently developed and deployed units. Unlike disappearing
frameworks and islands-based architectures, micro-frontends primarily address
organizational scalability and team autonomy rather than reducing JavaScript
execution in the browser. Nevertheless, they illustrate the broader effort to
manage the growing complexity of modern web applications.

While disappearing frameworks, Islands Architecture, ISR, and related approaches
differ in their implementation strategies, they share a common objective: reducing
the complexity, performance costs, and maintenance burden associated with modern
web applications. These approaches primarily focus on improving how applications
are rendered, delivered, and executed, while preserving the rich interactive
experiences expected by users.

A different line of work approaches the problem from an architectural perspective
by revisiting the principles of the original Web. Rather than focusing on
rendering and hydration strategies, these approaches return to the REST
architectural style proposed by Fielding \cite{fielding2000architectural},
particularly the \emph{Hypermedia Constraint} (HATEOAS). Within this context,
\emph{Hypermedia Systems} \cite{gross2023hypermedia} rely on \emph{hypermedia
controls} \cite{gross2024hypermedia}, which define how clients transition between
application states through server-provided representations.

Recent developments in this area build directly on the formalization of hypermedia
controls proposed by Gross~\cite{gross2024hypermedia}, in which controls are
defined as elements that (1) respond to events, (2) issue HTTP requests, and (3)
integrate server responses into designated locations within the user interface. In
REST terms, these controls act as \textit{affordances}: they tell the client which
application transitions are currently available and how they can be activated
next. This is central to HATEOAS, where the server does not only return data, but
also represents the valid next actions through links, forms, buttons, and related
controls embedded in the response. Frameworks and libraries such as
htmx~\cite{htmx}, Turbo~\cite{turbo}, and Datastar~\cite{datastar} extend this
model by enriching \textit{hypermedia controls} with additional client-side
capabilities, including fine-grained partial page updates, declarative DOM
placement strategies, and reactive behavior
coordination~\cite{vepsalainen2026revisiting}. These systems maintain the server
as the primary source of application state and interaction logic while delegating
only minimal orchestration responsibilities to the client, thereby reducing
reliance on full client-side application frameworks.

The htmx example in Listing~\ref{lst-htmx-button} illustrates how the hypermedia
control model can be expressed directly through HTML attributes. The
\texttt{hx-trigger="click"} attribute specifies the event that activates the
control, while \texttt{hx-post="/clicked"} defines the HTTP \texttt{POST}
request issued to the \texttt{/clicked} endpoint. The
\texttt{hx-target="\#parent-div"} attribute identifies the location where the
server response is integrated. 
The \texttt{hx-swap} attribute provides several options for controlling how htmx
integrates the server response into the DOM. In this example, \texttt{outerHTML}
replaces the entire target element with the returned HTML fragment.
Listing~\ref{lst-htmx-insert-element} shows the
corresponding HTTP response containing the HTML fragment inserted into the
target element. Thus, the button encapsulates event handling, request dispatch,
and response integration without requiring explicit client-side scripting.

\begin{lstlisting}[
  language=html,
  caption={Example of an htmx button \protect\footnotemark.},
  label={lst-htmx-button},
]
<button hx-post="/clicked"
    hx-trigger="click"
    hx-target="#parent-div"
    hx-swap="outerHTML">
    Click Me! 
</button>
\end{lstlisting}

\footnotetext{Example from \textit{htmx} documentation~\cite{htmxdocs}.}

\begin{lstlisting}[
caption={Example HTTP response for an HTMX inner HTML update},
label={lst-htmx-insert-element}
]
HTTP/1.1 200 OK
Content-Type: text/html

<div id="parent-div">
    <p>Hello world!</p>
</div>
\end{lstlisting}

Datastar adopts a different API~\cite{datastardocs} style based on the concept of
\textit{actions}, which are embedded within JavaScript-like expressions assigned
to Datastar attributes. Actions represent triggered behaviors, typically initiated
by user interactions or lifecycle events, and are denoted by the \texttt{@}
prefix. These expressions declaratively specify the operations to be performed,
such as issuing HTTP requests or updating reactive state, while remaining embedded
within HTML attributes. Listing~\ref{lst-datastar-button} presents the equivalent
Datastar implementation of the HTMX example shown in
Listing~\ref{lst-htmx-button}.

\begin{lstlisting}[
  language=html,
  caption={Example of a Datastar button.},
  label={lst-datastar-button},
]
<button data-on:click="@post('/clicked')">
    Click Me!
</button>
\end{lstlisting}

Instead of specifying the target element directly in the button attributes,
Datastar follows a different approach in which the backend determines both the
resulting HTML fragment and the location where that fragment should be integrated
into the document. To support this model, Datastar defines a patching protocol on
top of Server-Sent Events (SSE)\footnote{Standardized browser mechanism for
receiving a continuous stream of server-generated events over a single HTTP
connection.}, using the \texttt{text/event-stream} MIME type. SSE represents each
message as a sequence of key–value fields in plain text. In this format, each
message begins with an \texttt{event:} field that specifies the type of operation
being sent by the server, followed by one or more \texttt{data:} fields that carry
the associated payload. Each \texttt{data:} line represents a fragment of the
message, and multiple lines are concatenated in order to reconstruct the full
structured payload on the client side. Listing~\ref{lst-datastar-patch-elements}
illustrates an example of an HTTP response that patches the HTML element identified by
\texttt{parent-div}. The event type is \texttt{datastar-patch-elements},
indicating that the server is instructing the client to update a portion of the
DOM. The subsequent \texttt{data: elements} entries define the HTML fragment to be applied, split
across multiple lines for streaming and formatting purposes. When combined, these
lines reconstruct the complete HTML fragment: \texttt{<div id="parent-div"><p>Hello world!<p></div>},
which represents the content to be inserted or used as a replacement for the
selected element.

\begin{lstlisting}[
caption={Example of HTTP response for Datastar element patching},
label={lst-datastar-patch-elements}
]
HTTP/1.1 200 OK
Content-Type: text/event-stream

event: datastar-patch-elements
data: elements <div id="parent-div">
data: elements     <p>Hello world!<p>
data: elements </div>
\end{lstlisting}

Datastar provides several backend SDKs for commonly used technologies such as
Java, Python, Go, and Kotlin, which are designed to simplify the generation of
Datastar-specific Server-Sent Events (SSE).


%% file: sections/sec03-signal-base-dsl.tex
\section{HTML Templating and Safe Reactive Signal Management}
\label{sec:signals-dsl}

\begin{figure}
  \centering
  \captionsetup{justification=centering}
  \includegraphics[width=0.95\linewidth]{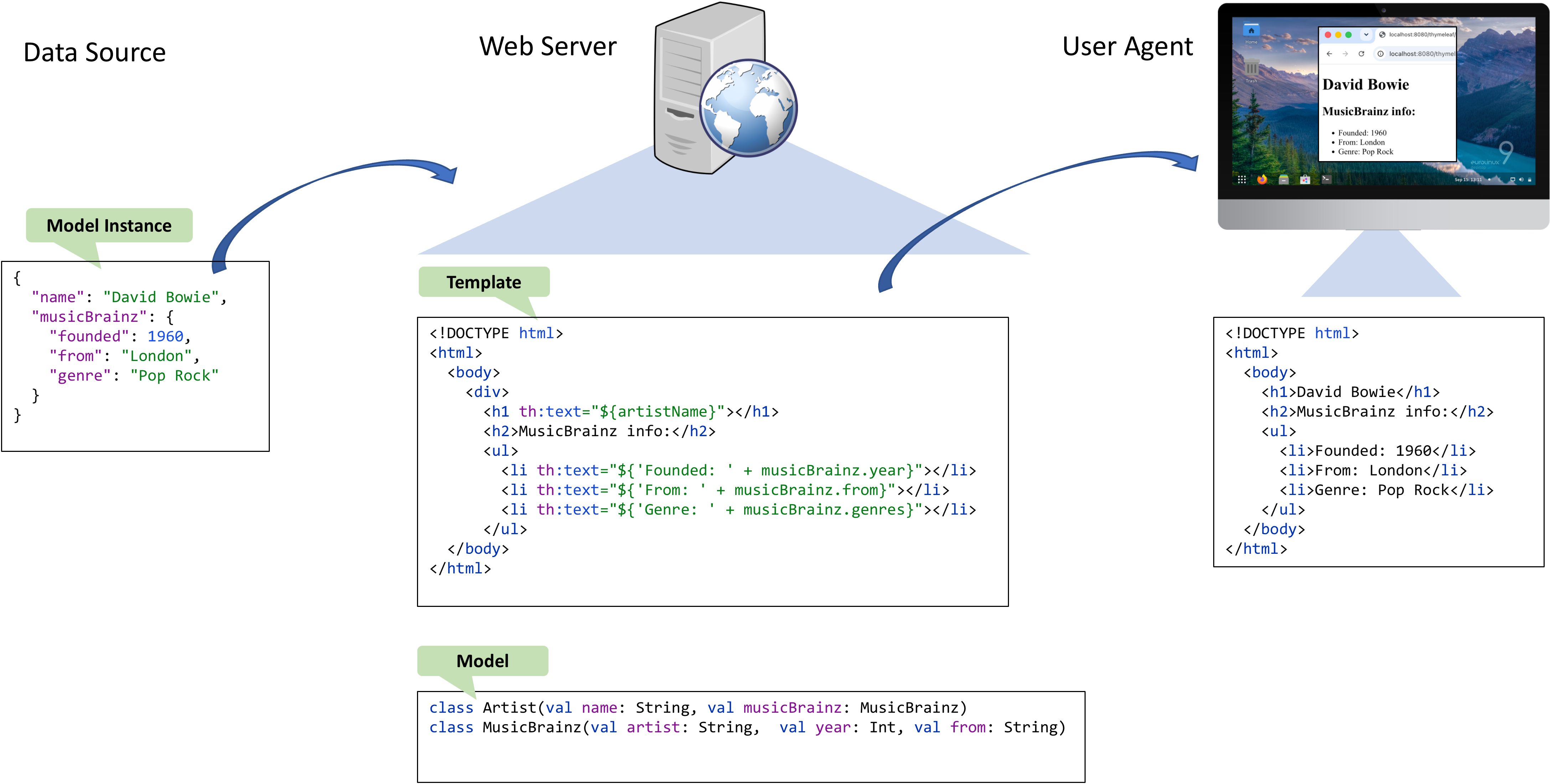}
  \caption{Example of server-side template rendering with a Thymeleaf template.}
  \label{sec03-template-workflow}
\end{figure}

This section establishes the architectural foundation for modern \textit{web templates},
drawing a distinction between \textit{external} and \textit{internal} templating idioms.
A \textit{web template} serves as a reusable layout for an HTML document that an
engine combines with \textit{input data models} to produce standard HTML as
output~\cite{Fowler02}.
Figure~\ref{sec03-template-workflow} illustrates a typical web templating
workflow using Thymeleaf. The template contains placeholders and expressions,
such as \texttt{th:text="\${artistName}"} and \texttt{th:text="\${'Founded: ' +
musicBrainz.year}"}, which define how dynamic content should be inserted into
the HTML document.
At runtime, the template is combined with an input data model containing
information about an artist, e.g. David Bowie and associated MusicBrainz data.
During the rendering process, Thymeleaf evaluates the template expressions
against the provided data model and replaces them with the corresponding values.
The resulting output is a complete HTML document in which all template
directives have been resolved.

Traditional templating approaches such as Thymeleaf, JSP, Handlebars, Pebble, or
FreeMarker embed dynamic expressions into HTML using string-based markup
extensions like \texttt{\textless\%}, \texttt{\{\{\}\}}, \texttt{\$\{\}}, which
lack native type safety and can introduce runtime
vulnerabilities~\cite{carvalho2020}. 
In contrast, an \textit{internal} Domain Specific Language (DSL) approach embeds
HTML templating directly into a statically compiled programming language,
providing structural verification at compile time. Building upon this paradigm,
this section outlines an approach that pairs statically typed templates with
backend-driven state updates.

This section is organized in three parts.
First, we describe the design and characteristics of an internal DSL for HTML,
focusing on how it leverages standard programming constructs to build user
interface components. Second, we formalize the signal-based reactive principles
that govern state propagation.
Finally, we detail how these two concepts converge to realize a signal-based,
statically typed DSL for HTML, enabling seamless and safe reactive hypermedia
controls.

\subsection{Internal Domain Specific Language (DSL) for HTML}

An \textit{internal} Domain Specific Language (DSL) like HtmlFlow, Hiccup,
ScalaTags, KotlinX.Html, or others, leverages its host programming language
(e.g., Java, Clojure, Scala, Kotlin) as the core dialect to define web
templates~\cite{carvalho2020}.
Although HtmlFlow was originally designed for Java, it also provides an
idiomatic Kotlin API.
Specifically HtmlFlow supports HTML builders using \textit{function
literals with receiver}~\cite{kotlinlang}.
In Kotlin, a block of code enclosed in curly braces \texttt{\{...\}} is known as
a \emph{lambda}, and can be used as an argument to
a function that expects a \emph{function literal}.
When we write, for example, \texttt{body \{ div \{ hr() \} \}}, we are
invoking the \texttt{body} function with a lambda as its argument. This
lambda, in turn, calls the \texttt{div} function with another lambda as an
argument that creates a horizontal row (i.e. \texttt{hr}).
Each call to an HTML builder (e.g., \texttt{body}, \texttt{div}, \texttt{hr})
creates the child element within the element generated by the outer function
call.

Listing~\ref{lst-htmlflow-basic} illustrates the HtmlFlow Kotlin idiom for
constructing an HTML template equivalent
to the Thymeleaf template presented in Figure~\ref{sec03-template-workflow}.
The provided \texttt{HtmlView} highlights a statically typed binding with the
\texttt{Artist} domain model.
Within this boundary, properties like \texttt{m.name} and the aggregated
\texttt{m.musicBrainz} fields (\texttt{year}, \texttt{from} and \texttt{genres})
are resolved dynamically yet checked at compile time, guaranteeing that data
bindings conform directly to the domain model.
Rather than parsing unverified text attributes, this view exclusively leverages
standard Kotlin programming constructs.

\begin{figure}
\begin{lstlisting}[
  language=kotlin,
  caption={Basic example of an idiomatic HtmlFlow Kotlin DSL template.},
  label={lst-htmlflow-basic},
]
val artistView: HtmlView<Artist> = view<Artist> {
    html {
        body {
            h1 { dyn { m: Artist ->  text(m.name) } }
            h2 { text("MusicBrainz info:") }
            ul { dyn { m: Artist -> with(m.musicBrainz) {
                li { text("Founded: " + year) }
                li { text("From: " + from) }
                li { text("Genre: " + genres) }
            }}}
        }
    }
}
\end{lstlisting}
\end{figure}

The \texttt{\textbf{dyn}} builder is employed to seamlessly integrate Kotlin
code into the definition of web templates and has the signature defined in
Listing~\ref{lst:dyn}.
It takes a function literal (i.e. \texttt{cons}) with a receiver corresponding
to the parent element (i.e. \texttt{T}) and an argument representing the model
(i.e. \texttt{M}). The \texttt{\textbf{dyn}} builder returns the same parent
element (i.e. \texttt{T}). 

\begin{lstlisting}[
  language=kotlin,
  basicstyle=\fontsize{8}{9}\ttfamily,
  label={lst:dyn},
  caption={HtmlFlow builder \texttt{dyn} to intertwine Kotlin constructs.},
  frame={tblr}
]
Element<T>.dyn(cons: T.(M) -> Unit): T
\end{lstlisting}

Given an HtmlFlow view, e.g. \texttt{artistView}, the model instance (e.g.
\texttt{bowie}) is later passed to its \texttt{write(model: T)} method to
produce the resulting HTML. In the following example, the view
\texttt{artistView} is first bound to an output stream \texttt{outStream}, and
then resolved with the \texttt{bowie} instance so that the rendered HTML is
written directly to it:
\texttt{artistView.setOut(outStream).write(bowie)}.

\subsection{Foundations of Reactive Signals}

Reactive signals originate from the broader class of \emph{reactive programming}
models~\cite{elliott1997functional}.
Signals are a programming abstraction used to model state that can be observed
and propagated through a system in a declarative manner. 
Early work on functional reactive programming (FRP) introduced the notion of
time-varying values and event-driven propagation as first-class abstractions for
structuring interactive systems~\cite{elliott1997functional,
frp_bainomugisha2013survey}. In these models, a signal is typically understood
as a continuous or discrete value that maintains dependencies on other signals,
enabling automatic re-evaluation when upstream data changes.

Formally, a signal can be seen as a function of time $S : t \rightarrow v$,
where changes in the underlying value $v$ trigger the re-evaluation of any
computation that depends on $S$. In contrast to traditional mutable variables,
signals introduce a dependency-tracking mechanism that builds a directed graph
of relationships between producers (state sources) and consumers (derived
computations or views). When a signal is updated, only the affected downstream
nodes in this dependency graph are recomputed, enabling efficient incremental
updates.

Computed signals define derived values as pure functions over one or more input
signals. Formally, a computed signal $C$ can be defined as $C(t) = f(S_1(t),
S_2(t), \dots, S_n(t))$, where $f$ is a pure function and $S_i$ are the input
signals on which $C$ depends. Any change in one of the input signals $S_i$
automatically triggers a recomputation of $C$, preserving consistency across the
dependency graph while maintaining referential transparency of derived state.

Modern reactive systems refine this idea by introducing \emph{fine-grained
reactivity}, where dependencies are tracked at the level of individual variables
or expressions rather than coarse execution units. Frameworks such as SolidJS
and similar signal-based architectures formalize signals as mutable cells with
explicit dependency tracking and minimal recomputation
overhead~\cite{carpenter2021solidjs}. This shift improves
performance by ensuring that only the affected portions of the computation graph
are updated when a signal changes.
This design direction is aligned with the emerging TC39 Signals
proposal~\cite{tc39signals}, which aims to standardize a minimal reactive
primitive for JavaScript. Although still under active specification, this
proposal reflects a convergence toward a unified model of fine-grained
reactivity across modern JavaScript runtimes.


\subsection{Signal-based Statically Typed DSL for HTML}

Datastar is a JavaScript library that implements a backend-driven hypermedia
approach.
Complementing Datastar attributes are \textit{actions} and \textit{signals}, which
together define the dynamic aspects of the system. To propagate state changes
efficiently, Datastar relies on a push-based notification protocol provided
through a modular plugin architecture. By default, Datastar uses 
Server-Sent Events (SSE\footnote{Server-Sent
Events (SSE) which is a unidirectional communication protocol that enables servers to
push real-time updates to clients over a standard HTTP connection. Compared with
WebSockets, SSE is simpler to implement when only server-to-client communication
is required.\cite{mdn-sse}})-based communication protocol that streams
incremental updates from the server to the client. Through this
\textit{unidirectional} channel, the server can issue targeted patch operations
that update both signal values and HTML elements without requiring a full page
reload.

The Datastar example of Listing~\ref{lst-html-counter} illustrates the
combination of \textit{signals} and \textit{actions} to implement a simple
reactive counter. 
Note that other optimized and recommended implementations of this example are
possible. However, this approach is intentionally used to illustrate the
complete end-to-end flow of signal propagation.
The root element declares a signal state using
\texttt{data-signals="{count: 0}"}, which initializes a reactive variable
\texttt{count} with the value 0. This signal is bound to the user interface
through the expression \texttt{data-text="\$count"} (line 4), ensuring that the
content of the \texttt{span} element is automatically updated whenever the
signal changes.
User interaction is handled through action-based event bindings on the buttons.
The attributes \texttt{data-on:click="@post('/decrement')"} and
\texttt{data-on:click="@post('/increment')"} define actions that are executed
when the user clicks the corresponding buttons. These actions dispatch a
\texttt{fetch} \texttt{POST} request to the specified path (\texttt{/decrement}
and \texttt{/increment}), triggering the corresponding server-side handler
responsible for updating the shared state.

\begin{lstlisting}[
  language=html,
  caption={A shared global counter maintain via Signals.},
  label={lst-html-counter},
]
<div data-signals="{count: 0}">
    <h1>Counting Stars - via Signals</h1>
    <div data-init="@get('/counter-signals/events')">
        <span id="counter" data-text="$count"></span>
    </div>
    <div>
        <button id="decrement" data-on:click="@post('/decrement')">-</button>
        <button id="increment" data-on:click="@post('/increment')">+</button>
    </div>
</div>
\end{lstlisting}

The reactive lifecycle is initialized using the \texttt{data-init} attribute,
which triggers a server request via the action
\texttt{@get('/counter-signals/events')}. This establishes a SSE connection
between client and server, enabling continuous backend-driven updates to be pushed
to the UI. Importantly, the counter state is shared across all connected frontend
clients, meaning that any update performed by one client is immediately reflected
in every other subscribed client through server-side propagation. When a button is
activated, the bound Hypermedia Control dispatches a \texttt{POST} request to the
corresponding endpoint. The server mutates its authoritative shared state and
immediately pushes the updated value to all connected clients using the
\texttt{datastar-patch-signals} event, as shown in
Listing~\ref{lst-counter-events}. This SSE payload represents a minimal patch
message where the event type \texttt{datastar-patch-signals} indicates a signal
update operation. The \texttt{data} field contains a partial state update,
specifying that the \texttt{count} signal should be updated to the value 1. Upon
receiving this event, each client reconciles the patch with its local signal
graph, resulting in an automatic update of all bound UI elements, such as the
\texttt{span} displaying \texttt{\$count}.

\begin{lstlisting}[
  caption={Server-emitted signal events},
  label={lst-counter-events}
]
event: datastar-patch-signals
data: signals { count: 1 }

\end{lstlisting}  

To support reactive \textit{hypermedia controls}, we extend the HtmlFlow DSL with
a signal-based API inspired by Reactive Signal models. In this approach, signals
are represented as statically typed symbolic references embedded directly into the
HTML DSL. Rather than behaving as locally reactive variables, signals act as typed
identifiers used to establish bindings between \textit{hypermedia controls},
request parameters, and backend-driven UI updates.

Listing~\ref{lst-signal-definition} presents the definition of the
\verb|Signal<T>| abstraction.
A signal encapsulates a typed symbolic reference associated with a name, where
the signal name must be unique within the template view definition to avoid
ambiguity in bindings and control resolution. The signal itself does not
maintain reactive state on the server runtime; instead, it provides a statically
typed representation that can be composed within Hypermedia Control expressions
and translated into the underlying Datastar-compatible syntax.

\begin{lstlisting}[
  language=kotlin,
  caption={Typed signal abstraction},
  label={lst-signal-definition},
]
class Signal<T>(val name: String)  {...}
\end{lstlisting}

Signals are introduced into HtmlFlow templates through the
\verb|dataSignal(...)| function, which creates typed signal bindings that can be
referenced throughout the HTML DSL. 
Listing~\ref{lst-counter-template} illustrates the equivalent HtmlFlow
definition of the example shown in Listing~\ref{lst-html-counter}.
Unlike the Datastar sample in Listing~\ref{lst-html-counter}, which relies on
string interpolation or inline JavaScript expressions, the proposed API enables
signal composition through native Kotlin expressions and typed builders that are
verified at compile time.
No string literals are used for signal binding or endpoint routing. Instead,
signal binding is performed through direct variable references (e.g.
\texttt{count} in line 7), ensuring that renaming the signal is consistently
reflected across the entire template at compile time. Similarly, server
interactions are not defined using string-based URL paths, but through function
references using Kotlin’s \texttt{::} operator, which binds directly to backend
route handlers within the project scope. In this model, \texttt{get} and
\texttt{post} are first-class functions rather than string-encoded HTTP verbs,
further reinforcing compile-time safety and structural consistency between
template and server logic.

\begin{lstlisting}[
  language=kotlin,
  caption={Server-backed signal (\texttt{count}) drives reactive UI updates.},
  label={lst-counter-template}
]  
div {
    val count: Signal = dataSignal("count", 0)
    div {
      dataInit { get(::getCounterEventsSignals) }
      span {
        attrId("counter")
        dataText{ +count }
      }
    }
    div {
      button { dataOn(Click) { post(::decrementCounterViaSignals) }}
      button { dataOn(Click) { post(::incrementCounterViaSignals) }}
    }
}
\end{lstlisting}

This approach allows Hypermedia Control definitions, signal bindings, and HTML
structure to coexist within a single homogeneous programming model.
By embedding signals directly into the HtmlFlow DSL, the proposed model improves
type safety, reduces reliance on string-based expressions, and enables
declarative backend-driven reactivity without requiring application-specific
client-side JavaScript.

%% file: sections/sec04-patterns.tex
\section{Backend-driven Reactive Hypermedia Control Patterns}
\label{sec:patterns}

This section presents a set of Hypermedia Control patterns based on a
backend-driven reactive model. In this approach, some controls are bound to signals
that represent application state, and changes in these signals may trigger control
reactions. A control reacts to a signal change by dispatching an HTTP request,
whose response may update the same signal and/or partially patch the corresponding
HTML region.
This establishes a bidirectional loop in which signals drive UI behavior, while
the server responses continuously refine both state and presentation. The
resulting model unifies state management and interaction through hypermedia,
reducing the need for client-side imperative logic and enabling declarative,
backend-driven reactivity at the level of individual controls.
The following patterns illustrate how this reactive model can be applied to
different interaction scenarios:

\begin{enumerate}
  \item Backend-Driven Signal Update
  \item Action-Triggered HTML Patch
  \item Signal-Based HTML Patch
  \item Signal-Based Incremental HTML Patch
  \item Signal-based Formless Submission
\end{enumerate}

Together, the Backend-driven Reactive Hypermedia Control (BRHC) patterns 
cover a range of backend-driven
behaviors, from simple state synchronization mechanisms to more advanced
interactions that combine signal propagation, partial HTML updates, and
Hypermedia Control-driven requests.

\subsection{Backend-Driven Signal Update}
\label{counter-pattern}

The first pattern illustrates how reactive UI behavior can be coordinated
through backend-driven signal updates. Instead of polling or issuing explicit
refresh requests, the client establishes a persistent SSE connection through
which the server pushes signal updates whenever application state changes.
This pattern is illustrated in Listing~\ref{lst-counter-template}, where the
signal named \texttt{count} is initialized with the value 0. The
\texttt{dataInit} attribute triggers the browser to open a persistent
SSE connection to \texttt{getCounterEventsSignals} as soon as
the element is mounted. The \texttt{dataText} binding ensures that the
\texttt{span} element is automatically re-rendered whenever \texttt{count}
changes.
The bound UI fragment is updated transparently without requiring explicit
client-side DOM manipulation or imperative rendering logic. The signal therefore
acts as a reactive synchronization mechanism between the authoritative server
state and the rendered UI.

Listing~\ref{lst-counter-ktor-sse} presents the backend route responsible for
maintaining a persistent connection and emitting signal updates. 
The \texttt{MutableStateFlow}~\cite{state-flow-kotlin} (line 1) is initialized with an
initial value of 0 and acts as a hot, observable
stream~\cite{kleppmann2017designing} that holds the current authoritative value
of the counter while immediately propagating any state updates to all active
collectors.
The route defined by \texttt{getCounterEvents} responds with a
\texttt{text/event-stream} content type (line 5) and suspends execution while collecting
values from a shared \texttt{MutableStateFlow} representing the counter
state (line 6).
Each time the \texttt{counter} flow is updated, the backend resumes execution of
\texttt{getCounterEvents} and invokes \texttt{patchSignals}, which serializes
the updated state and writes it to the response channel as a Server-Sent Event.
This ensures that all
connected clients remain synchronized with the latest counter value without
requiring polling, explicit refresh requests, or client-side state
reconciliation logic.

\begin{lstlisting}[
  language=kotlin,
  caption={Backend route responsible for streaming signal updates to connected clients.},
  label={lst-counter-ktor-sse}
]
private val counter: MutableStateFlow<Int> = MutableStateFlow(0)

private suspend fun RoutingContext.getCounterEvents() {
    call.respondBytesWriter(ContentType.Text.EventStream, OK) {
        with(ServerSentEventGenerator(response(this))) {
            counter.collect {
                patchSignals("{count: ${counter.value}}")
            }
        }
    }
}
\end{lstlisting}

\begin{figure}
  \centering
  \captionsetup{justification=centering}
  \includegraphics[width=0.65\linewidth]{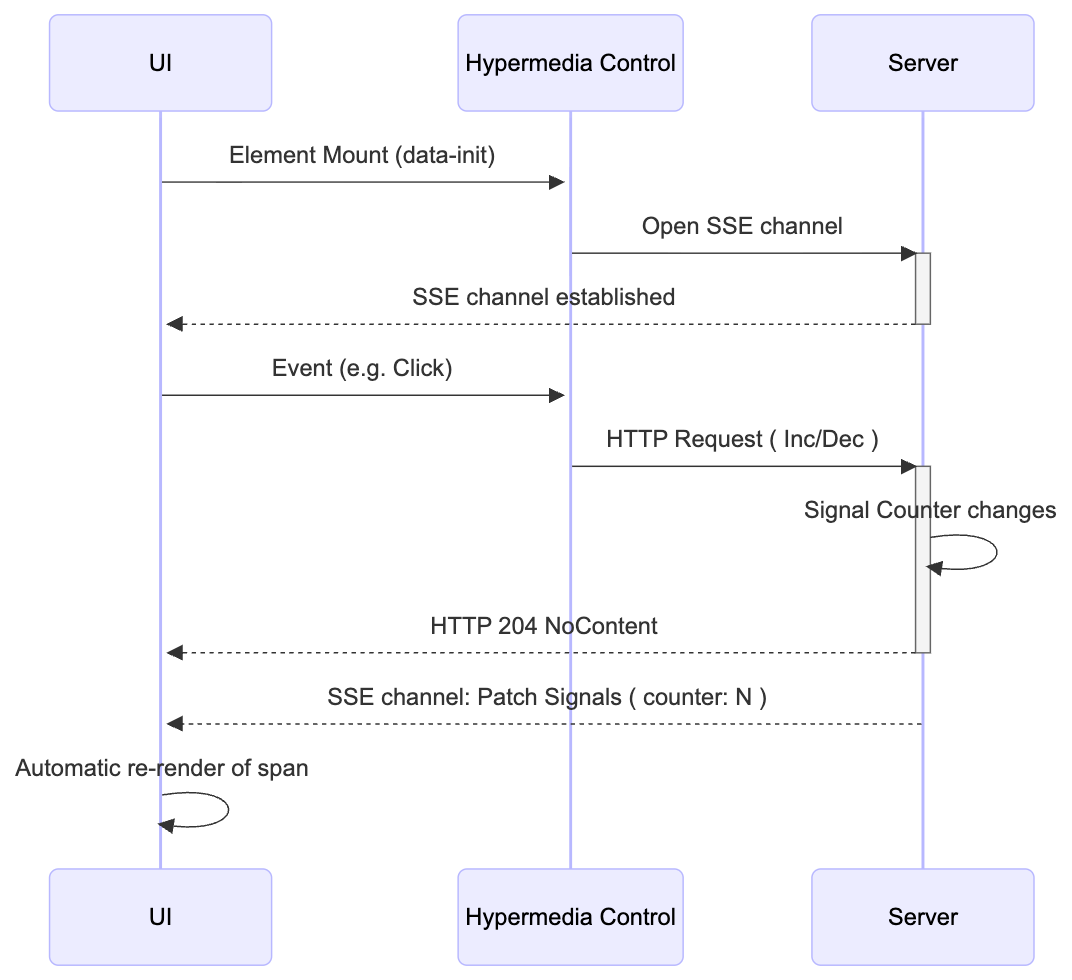}
  \caption{Backend-driven reactive loop with signal update.}
  \label{counter-seq-diagram}
\end{figure}

Figure~\ref{counter-seq-diagram} summarizes the interaction flow of the
Backend-Driven Signal Update pattern described previously. The diagram highlights
the main participants involved in the interaction: the user interface, the
Hypermedia Control, and the server. In this context, a Hypermedia Control
corresponds to an HTML element enhanced with a Datastar attribute capable of
triggering HTTP interactions.

\subsection{Action-Triggered HTML Patch}
\label{dsl-pattern2}

This pattern illustrates a Hypermedia Control that triggers a backend-driven UI
update in response to an action. When the Hypermedia Control is activated, the
client does not determine how the interface should change. Instead, it
dispatches the HTTP request defined by the action, while the backend processes
the request and determines the corresponding UI update.

\begin{figure}
  \centering
  \captionsetup{justification=centering}
  \includegraphics[width=0.45\linewidth]{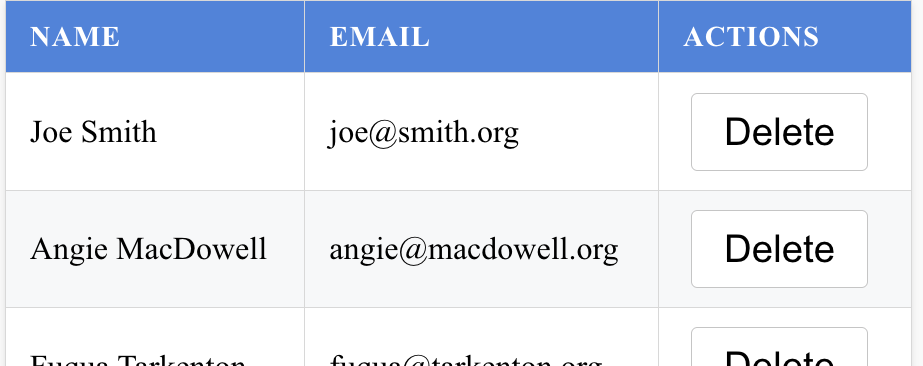}
  \caption{Delete row example applying Action-based Backend-driven pattern.}
  \label{delete-row-table}
\end{figure}

To illustrate a use case of this pattern, we use an example of a table that
includes a delete button on each row, allowing the corresponding table row to be
removed, as illustrated in Figure~\ref{delete-row-table}. 
The function \texttt{hfTableRow} presented in Listing~\ref{lst-delete-row} is an
extension over a table body element and generates a \texttt{tr} element whose
identifier is dynamically assigned using the index parameter (i.e.,
\texttt{row-\$index}), enabling precise targeting of individual rows. Each row
contains two data cells that display the user's name and email address,
respectively, followed by a third cell that defines an interactive control. The
\texttt{button} element is bound to a backend-driven hypermedia action through
\texttt{dataOn(Click)}, which issues an HTTP \texttt{DELETE} request to the path
\texttt{/delete-row/\$index} when activated, mutating the authoritative state,
and determining how the UI should be updated afterward.

\begin{figure}
\begin{lstlisting}[
  language=kotlin,
  caption={Action-based Backend-driven Row Deletion},
  label={lst-delete-row}
]
fun Tbody<*>.hfTableRow(index: Int, user: TableUser) = tr {
    attrId("row-$index")
    td { text(user.name) }
    td { text(user.email) }
    td {
        button {
            dataOn(Click) { delete("/delete-row/$index") }
            text("Delete")
        }
    }
}
\end{lstlisting}
\end{figure}

Listing~\ref{lst-delete-row-server-endpoint} presents the backend route
responsible for handling the deletion request. When the client activates the
delete control, the endpoint responds with a \texttt{text/event-stream} payload
containing a Datastar element patch operation.
The \texttt{PatchElementsOptions} object specifies both the target element and
the patching strategy. In this example, the CSS selector \texttt{\#row-\$rowIndex}
identifies the table row to be modified, while the
\texttt{ElementPatchMode.Remove} mode instructs Datastar client to remove the
matching element from the DOM. Upon receiving the corresponding SSE event,
Datastar locates the element with the identifier given by \texttt{row-\$rowIndex} and removes it
from the rendered page.
The example in Listing~\ref{lst-delete-row-patch-elements} shows a server
response containing an element patch operation that instructs the client to
remove the element identified, for instance, by \texttt{row-2}.

\begin{lstlisting}[
  language=kotlin,
  caption={Backend route responsible for deleting row via HTML element patch},
  label={lst-delete-row-server-endpoint}
]
suspend fun RoutingContext.deleteRow() {
    call.respondBytesWriter(ContentType.Text.EventStream, OK) {
        with(ServerSentEventGenerator(response(this))) {
            val rowIndex = call.pathParameters["index"]
            removeUser(rowIndex)
            patchElements(
                PatchElementsOptions("#row-$rowIndex", ElementPatchMode.Remove)
            )
      }
    }
}
\end{lstlisting}

\begin{lstlisting}[
  caption={Server-emitted HTML patch},
  label={lst-delete-row-patch-elements}
]
event: datastar-patch-elements
data: selector #row-2
data: mode remove
\end{lstlisting}  

\begin{figure}
  \centering
  \captionsetup{justification=centering}
  \includegraphics[width=0.50\linewidth]{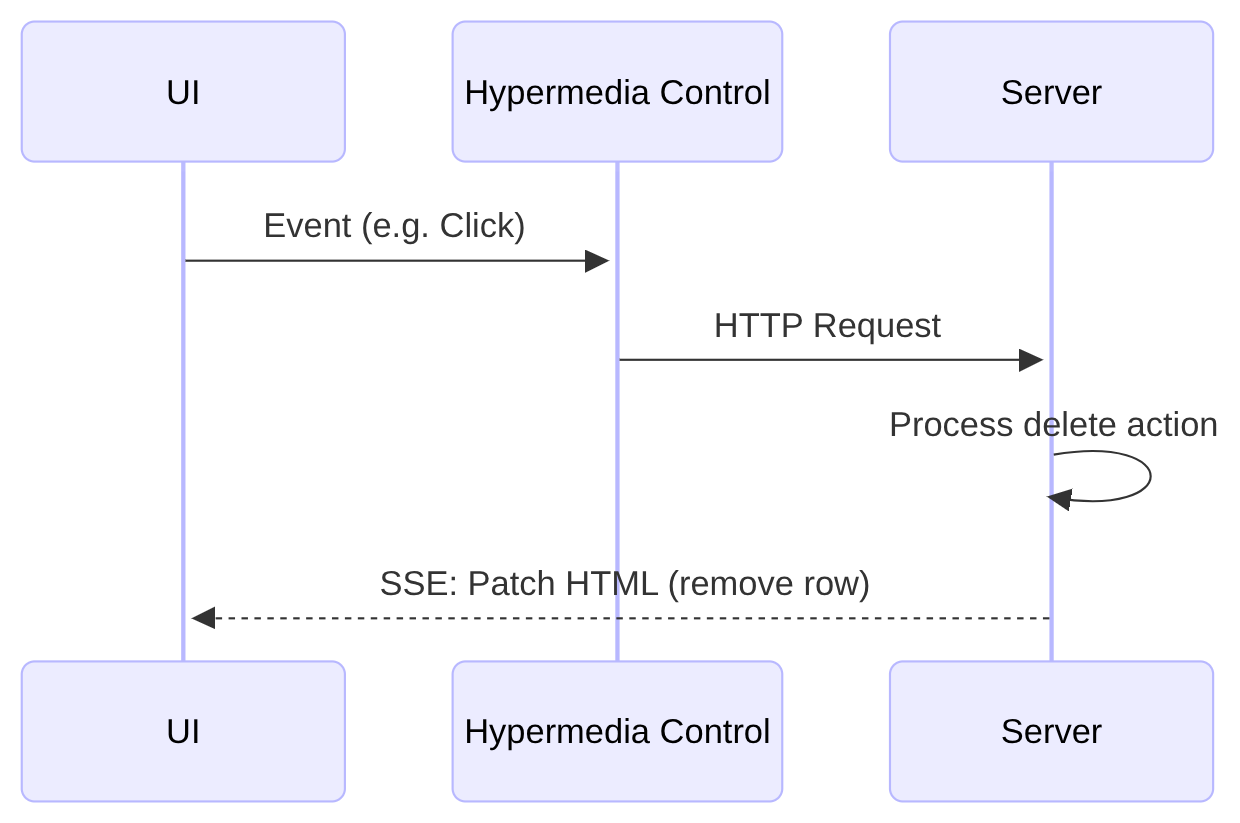}
  \caption{Action-based backend-driven HTML patch.}
  \label{delete-row-seq-diagram}
\end{figure}

Compared with the previous patterns, this one does not rely on signals as state
carriers as shown in Figure~\ref{delete-row-seq-diagram}. It establishes a
baseline in which a Hypermedia Control supports interactive behavior through a
direct action-response cycle alone. The control dispatches a request, and the
server response patches the affected HTML region.

\subsection{Signal-Based HTML Patch}
\label{dsl-pattern3}

This pattern extends \textit{hypermedia controls} with reactive user input. Rather
than relying solely on discrete user actions such as button clicks, the control
reacts to continuous input events and uses the resulting client-side state to
parameterize server requests. In this case, the state is represented by a search
term stored in a reactive signal, which is automatically included in outgoing
requests whenever the control is triggered. The responsibility for evaluating the
search criteria and generating the updated user interface remains entirely on the
server. The client only captures user input and declares the reactive behavior,
while the server performs the filtering operation and produces the corresponding
HTML representation.

\begin{figure}[h]
  \centering
  \captionsetup{justification=centering}
    \includegraphics[width=0.95\linewidth]{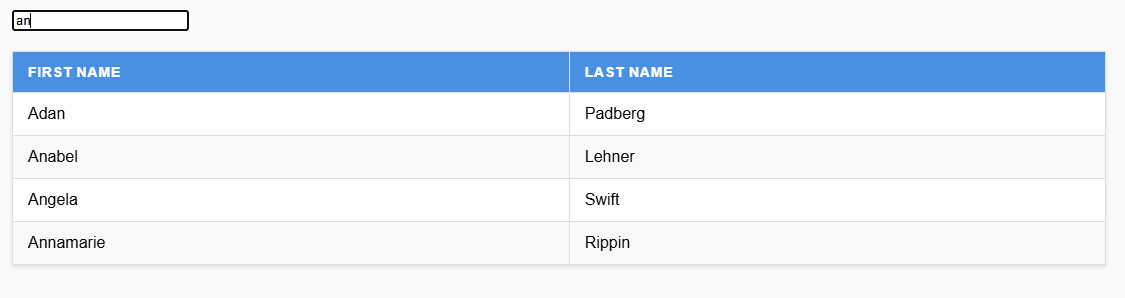}
  \caption{Reactive contact filtering user interface}
  \label{active-search-table-img}
\end{figure}

To illustrate this pattern, we use an active-search interface composed of a
contact table and a search input field. As the user types a search term, the
current value is synchronized through a signal and used to retrieve the matching
contacts from the server. The resulting interface is shown in
Figure~\ref{active-search-table-img}.

Listing~\ref{lst-activeSearch-template} presents the HtmlFlow code for the
reactive contact filtering interface, consisting of a search input and a results
table. The \texttt{input} element defines an event-driven Hypermedia Control that
dispatches an HTTP request whenever the \texttt{Input} event is triggered and no
subsequent input occurs within a debounce interval of 200 milliseconds. This
mechanism ensures that requests are issued only after the user pauses typing,
reducing unnecessary server interactions.

The \texttt{dataBind} attribute binds the current value of the input field to the
\texttt{search} signal.
This signal is automatically included in outgoing requests, such as the HTTP
\texttt{GET} request issued to the backend route handler
\texttt{getSearchContacts}, and represents the current query state of the user
interface. Rather than directly
initiating requests, the signal functions as a shared state carrier between client
and server, allowing the server to process the most recent user input when the
control is activated.

\begin{figure}
\begin{lstlisting}[
  language=kotlin,
  caption={Input-triggered hypermedia control for reactive contact filtering},
  label={lst-activeSearch-template}
]  
div {
    input {
        val search: Signal  = dataBind("search")
        dataOn(Input) {
            get(::getSearchContacts)
            modifiers { debounce(200.milliseconds) }
        }
    }
    table {
        thead { ... }
        tbody {
            attrId("contacts")
            ...
        }
    }
}
\end{lstlisting}
\end{figure}

Listing~\ref{lst-activeSearch-ktor} presents the backend route responsible for
handling search requests. When a Hypermedia Control triggers an HTTP request, it
automatically includes a query parameter named \texttt{datastar}, whose value is
a JSON representation of all currently defined signals, excluding local signals
whose names begin with an underscore (this regular expression can be modified through the Datastar configuration).
In turn, the endpoint responds with a \texttt{text/event-stream} payload
containing an HTML patch operation.
It first retrieves the \texttt{datastar} query parameter (line 6) and decodes
the \texttt{search} signal (line 7) using the \texttt{ActSearchSignals} data
class. 
After that, it processes the search query by invoking
\texttt{getContacts(search)} to retrieve the matching contacts and then renders
the resulting table body using the reusable HtmlFlow fragment
\texttt{activeSearchTbody} (line 8).
The generated HTML is stored in \texttt{tableBodyFragment}, which is
subsequently passed to \texttt{patchElements} (line 9). This operation serializes the
rendered fragment into a Datastar element patch and streams it to the client
through the SSE connection, causing the corresponding portion of the page to be
updated with the filtered search results.
An example of the emitted patch event is shown
in Listing~\ref{lst-activeSearch-html-event}.

\begin{lstlisting}[
  language=kotlin,
  caption={Backend route responsible for streaming HTML fragments to connected clients.},
  numbers=left,
  label={lst-activeSearch-ktor}
]
data class ActSearchSignals(val search: String)

private suspend fun RoutingContext.searchContacts() {
    call.respondBytesWriter(ContentType.Text.EventStream, OK) {
        with(ServerSentEventGenerator(response(this))){
            val dsQueryArg = call.request.queryParameters["datastar"]
            val (search) = Json.decodeFromString<ActSearchSignals>(dsQueryArg)
            val tableBodyFragment = activeSearchTbody.render(getContacts(search))
            patchElements(tableBodyFragment)
        }
    }
}
\end{lstlisting}  

\begin{lstlisting}[
  caption={Server-emitted HTML patch event.},
  label={lst-activeSearch-html-event}
]
event: datastar-patch-elements
data: elements <tbody id="contacts">
data: elements <tr><td> ... </td></tr>
\end{lstlisting}  

The generated HTML fragment contains a \texttt{tbody} element with the identifier
\texttt{contacts}, allowing Datastar's morphing mechanism to match it against the
corresponding element already present in the DOM. This identifier is therefore
required for the patch operation to be applied correctly. Upon receiving the
event, the client replaces the contents of the existing \texttt{\#contacts} table
body with the newly generated rows. 

\begin{figure}[h]
  \centering
  \captionsetup{justification=centering}
    \includegraphics[width=0.60\linewidth]{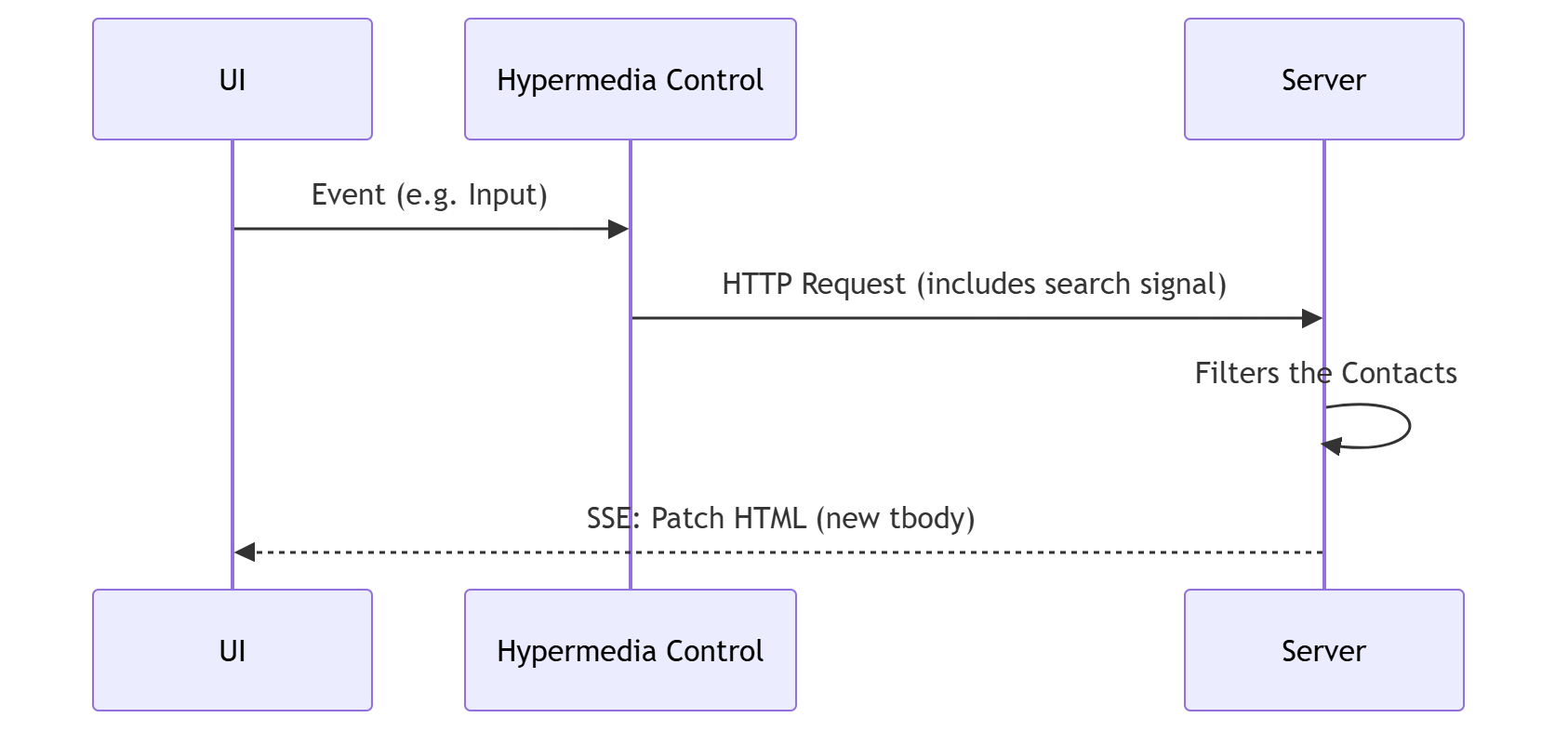}
  \caption{Backend-driven HTML patch with reactive input control.}
  \label{active-search-seq-diagram}
\end{figure}

Compared with the first pattern, where the server is responsible solely for
emitting signal updates while the client handles the corresponding UI updates,
this pattern shifts the responsibility of rendering back to the server. Instead of
relying on client-side signal reconciliation, the server directly produces the
HTML updates required to reflect the new state
(Figure~\ref{active-search-seq-diagram}). 

\subsection{Signal-Based Incremental HTML Patch}
\label{dsl-pattern4}

This pattern combines two complementary mechanisms: backend-driven signal updates
and partial HTML patching. In this approach, the client maintains a small reactive
state through signals, which are continuously updated by the server and later
reused by \textit{hypermedia controls} to drive subsequent interactions. Although
signals are stored on the client, the overall model remains backend-driven because
both the signal values and the HTML patches are generated and controlled by the
server. The client only acts as a temporary holder of interaction state required
to parameterize future requests.

To illustrate this pattern, we use an infinite-scroll interaction model in which a
table is progressively populated as the user scrolls toward the bottom of the
page. Listing~\ref{lst-intersection-template} presents the code for this model.
Two signals, \texttt{offset} and \texttt{limit}, are initialized through
\texttt{dataSignals} and represent the current pagination state. The initial table
body is populated with the first ten rows, and a sentinel div element at the
bottom of the page carries the Hypermedia Control that drives subsequent data
loading.
The Hypermedia Control is defined through the \texttt{dataOnIntersect} attribute.
When the associated element enters the viewport, the control automatically
dispatches an HTTP request to the server. The request implicitly includes the
current signal values, allowing the server to determine which additional rows must
be retrieved and appended to the table.

\begin{figure}
\begin{lstlisting}[
  language=kotlin,
  caption={Infinite-scroll Hypermedia Control using signal-driven pagination},
  label={lst-intersection-template}
]  
div {
    dataSignals("offset" to 10, "limit" to 5)
    table {
        thead {
            tr { th { text("Name") }; th { text("Email") }; th { text("ID") }}
        }
        tbody {
            attrId("agents")
            // First 10 rows (Agent Smith 0-9)
            for (i in 0..9) {
                tr { /* Agents Content */ }
            }
        }
    }
}
div { dataOnIntersect { get(::getMoreAgents) } }
\end{lstlisting}
\end{figure}

Listing~\ref{lst-infinite-scroll-server-code} presents the backend route
responsible for implementing the infinite-scroll interaction.

\begin{figure}
\begin{lstlisting}[
  language=kotlin,
  caption={Backend route for signal updates and incremental HTML patching.},
  label={lst-infinite-scroll-server-code}
]
data class Signals(val offset: Int, val limit: Int)

suspend fun RoutingContext.getMoreAgents() {
    call.respondBytesWriter(ContentType.Text.EventStream, OK) {
        with(ServerSentEventGenerator(response(this))) {
            val datastarArg = call.request.queryParameters["datastar"]

            // Decode the signals from the datastar query argument
            // and update the signals for the next request
            val (offset, limit) = Json.decodeFromString<Signals>(datastarArg)
            patchSignals("{offset: ${offset + limit}}")

            // Generate the new rows to be added to the table
            // and send the patch to the client
            val agents = newAgents(offset, offset + limit)
            val htmlRows = generateRows(agents)
            patchElements(htmlRows, PatchElementsOptions("#agents", Append))
        }
    }
}
\end{lstlisting}  
\end{figure}

The route shown in Listing~\ref{lst-infinite-scroll-server-code}
receives the current client signals through the \texttt{datastar} query parameter.
The parameter contains a JSON representation of the \texttt{offset} and
\texttt{limit} signals, which are deserialized into a \texttt{Signals} instance.
After obtaining the current pagination state, it updates the
\texttt{offset} signal by adding the current \texttt{limit} value and emits the
new signal value through \texttt{patchSignals}, which emits a
\texttt{datastar-patch-signals} SSE event, as illustrated in
Listing~\ref{lst-infinite-scroll-signals-event}.
This ensures that subsequent requests will continue from the correct position in
the dataset. After that, it retrieves the next batch of agents by invoking
\texttt{newAgents(offset, offset + limit)} and generates the corresponding HTML
table rows through \texttt{generateRows}.
Finally, the generated rows are streamed back to the client using
\texttt{patchElements}, which emits a \texttt{datastar-patch-elements} event, as
shown in Listing~\ref{lst-infinite-scroll-HTML-event}.
The patch operation targets the element identified by the CSS selector
\texttt{\#agents} and uses the \texttt{Append} mode, causing the newly generated
rows to be appended to the existing table contents rather than replacing them.

\begin{lstlisting}[
  caption={Datastar SSE event updating the \texttt{offset} signal.},
  label={lst-infinite-scroll-signals-event},
]
event: datastar-patch-signals
data: signals { offset: 15 }
\end{lstlisting}  

\begin{lstlisting}[
  caption={Server-emitted HTML events},
  label={lst-infinite-scroll-HTML-event},
]
event: datastar-patch-elements
data: selector #agents
data: mode append
data: elements <tr><td> Agent Smith 10 </td><td> ... </td><td> ... </td></tr>
...
data: elements <tr><td> Agent Smith 14 </td><td> ... </td><td> ... </td></tr>
\end{lstlisting}

\begin{figure}[h]
  \centering
  \captionsetup{justification=centering}
    \includegraphics[width=0.55\linewidth]{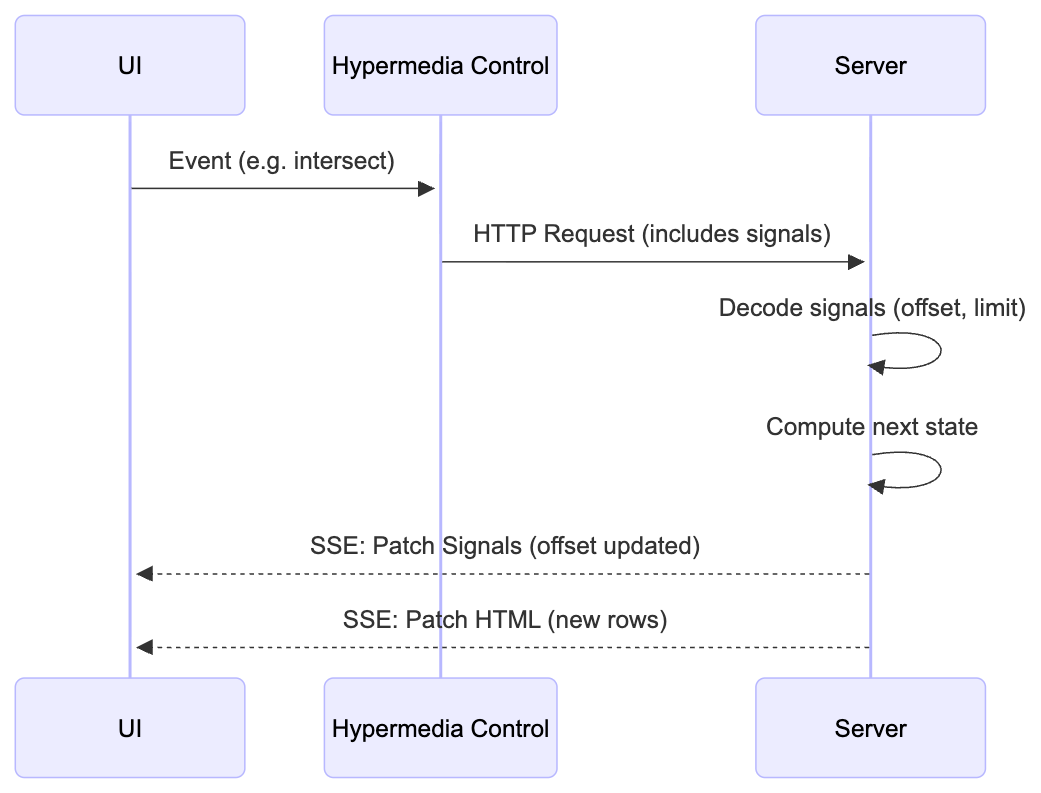}
  \caption{Backend-driven reactive loop with signal and HTML patching.}
  \label{infinite-load-seq-diagram}
\end{figure}

This interaction establishes a reactive feedback loop in which server-updated
signals continuously drive future Hypermedia Control interactions while the HTML
interface is progressively refined through partial patches. 
Compared with the Backend-Driven HTML Patch pattern described in Section 4.3,
where the server directly replaces a rendered region in response to a discrete
user action, this pattern introduces client-side signal state that persists across
multiple interactions. Rather than discarding intermediate state after each
response, the client retains the updated signal values and reuses them to
parameterize the next request. 
The server therefore does not need to embed pagination context within the HTML
itself; instead, it delegates that responsibility to the signal store, keeping
both the HTML fragments and the backend routes free of session-specific state.
Figure~\ref{infinite-load-seq-diagram} summarizes the full interaction flow of
this pattern.

\subsection{Signal-based Formless Submission}


Traditionally, data submission is performed through HTML forms, which act as
containers responsible for collecting and serializing input values before sending
them to the server. In this pattern, that responsibility is delegated to reactive
signals. Input elements are directly bound to signals, and \textit{hypermedia
controls} automatically include the current signal values when dispatching
requests. As a result, form-like interactions can be implemented without relying
on explicit \texttt{form} elements or manual client-side serialization logic.

\begin{figure}[h]
  \centering
  \captionsetup{justification=centering}
  \includegraphics[width=0.50\linewidth]{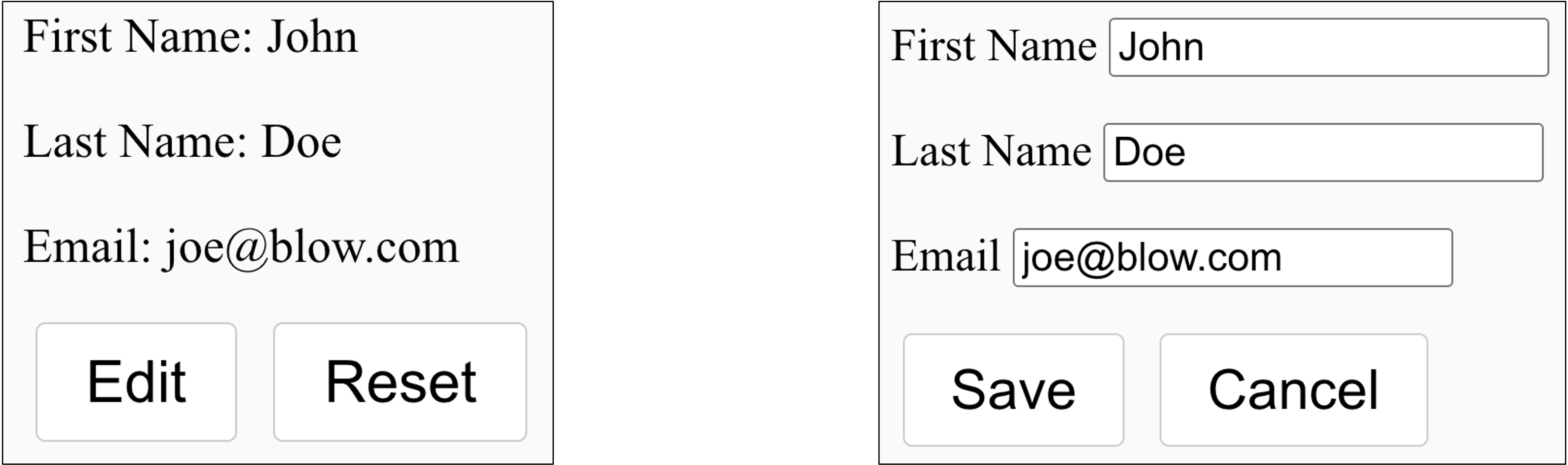}
  \caption{Read-only and edit views for the the Signal-based Formless Submission pattern.}
  \label{formless-submission-view-mode}
\end{figure}

Figure~\ref{formless-submission-view-mode} illustrates the two interface states
involved in the interaction. The first corresponds to a read-only view of the user
data, while the second exposes editable input fields bound to reactive signals.
The server controls transitions between these states by emitting HTML patch
operations that replace the current representation with the appropriate one.

Listing~\ref{lst-click-to-edit-viewMode} presents the HtmlFlow code responsible
for the read-only representation. This view displays the current contact
information and provides two \textit{hypermedia controls} that allow the user to
transition to a different presentation state. The \texttt{Edit} control requests
the editable representation, while the \texttt{Reset} control restores the
original contact information. In both cases, the server responds with an HTML
patch that replaces the corresponding interface region with the appropriate
representation, without the need of a full page reloads.

\begin{figure}
\begin{lstlisting}[
  language=kotlin,
  caption={Reactive read-only view synchronized HTML Patch},
  label={lst-click-to-edit-viewMode}
]
div { attrId("edit-contact")
    p { text("First Name: John") }
    p { text("Last Name: Doe") }
    p { text("Email: joe@blow.com") }
    div {
        button { 
            dataOn(Click) { get(::editContact) }
            text("Edit")
        }
        button { 
            dataOn(Click) { patch(::clickToEditSignalsReset) }
            text("Reset")
        }
    }
}  
\end{lstlisting}  
\end{figure}

Listing~\ref{lst-click-to-edit-endpoint-edit} presents the backend route
responsible for handling the edit request. The endpoint, retrieves the contact
information from a data acess layer and then render the HTML
which remains the authoritative application state. After persisting the updated
values, the endpoint responds with a \texttt{text/event-stream} payload containing
an HTML patch of the read-only representation. This patch replaces the element
identified by \texttt{edit-contact}, allowing the interface to transition back to
the read-only view while reflecting the updated contact information.

\begin{lstlisting}[
  language=kotlin,
  caption={Backend route responsible for received the Edit request.},
  label={lst-click-to-edit-endpoint-edit}
]
private suspend fun RoutingContext.editContact() {
    call.respondBytesWriter(ContentType.Text.EventStream, OK) {
        with(ServerSentEventGenerator(response(this))) {
            val contactInfo = getContact()
            patchElements(hfReadOnlyFragment.render(contactInfo))
        }
    }
}
\end{lstlisting}  

The editable representation, shown in Listing~\ref{lst-click-to-edit-editMode},
presents the HtmlFlow code responsible for editing the contact information. The
view contains three input elements, each bound to a corresponding reactive signal
through \texttt{dataBind}. When the \texttt{Save} control is activated, the
\textit{hypermedia control} dispatches an HTTP request. The current values of all bound
signals are automatically serialized and included in the request body, without
requiring an explicit HTML form or imperative JavaScript serialization logic.

\begin{lstlisting}[
  language=kotlin,
  caption={Signal-based formless submission through reactive input bindings},
  label={lst-click-to-edit-editMode},
]
div {
    attrId("edit-contact")
    p { text("First Name"); input { dataBind(firstName) } }
    p { text("Last Name"); input { dataBind(lastName) } }
    p { text("Email"); input { dataBind(email) } }
    div {
        button {
            dataOn(Click) { put(::saveClickToEdit) }
            text("Save")
          }
        button {
            dataOn(Click) { get(::cancelClickToEdit)}
            text("Cancel")
          }
    }
}
\end{lstlisting}  

Listing~\ref{lst-click-to-edit-endpoint-save} presents the backend route
responsible for handling the save request. The endpoint decodes the serialized
signals received in the request body, updates the server-side contact
information, and responds with a \texttt{text/event-stream} payload containing
an HTML patch of the read-only representation. This patch replaces the element
identified by \texttt{edit-contact}, causing the interface to return to the
read-only view with the updated contact information.

\begin{figure}
\begin{lstlisting}[
  language=kotlin,
  caption={Backend route responsible for received the Save request.},
  label={lst-click-to-edit-endpoint-save}
]
class Signals(val first: String, val last: String, val email: String)

private suspend fun RoutingContext.saveClickToEdit() {
    call.respondBytesWriter(ContentType.Text.EventStream, OK) {
        with(ServerSentEventGenerator(response(this))) {
            val datastarBodyArgs = call.request.call.receiveText()
            val updatedSignals = Json.decodeFromString<Signals>(datastarBodyArgs)
            saveNewContactInfo(updatedSignals)
            patchElements(hfReadOnlyFragment.render(updatedSignals))
        }
    }
}
\end{lstlisting}
\end{figure}

\begin{figure}[h]
  \centering
  \captionsetup{justification=centering}
    \includegraphics[width=0.55\linewidth]{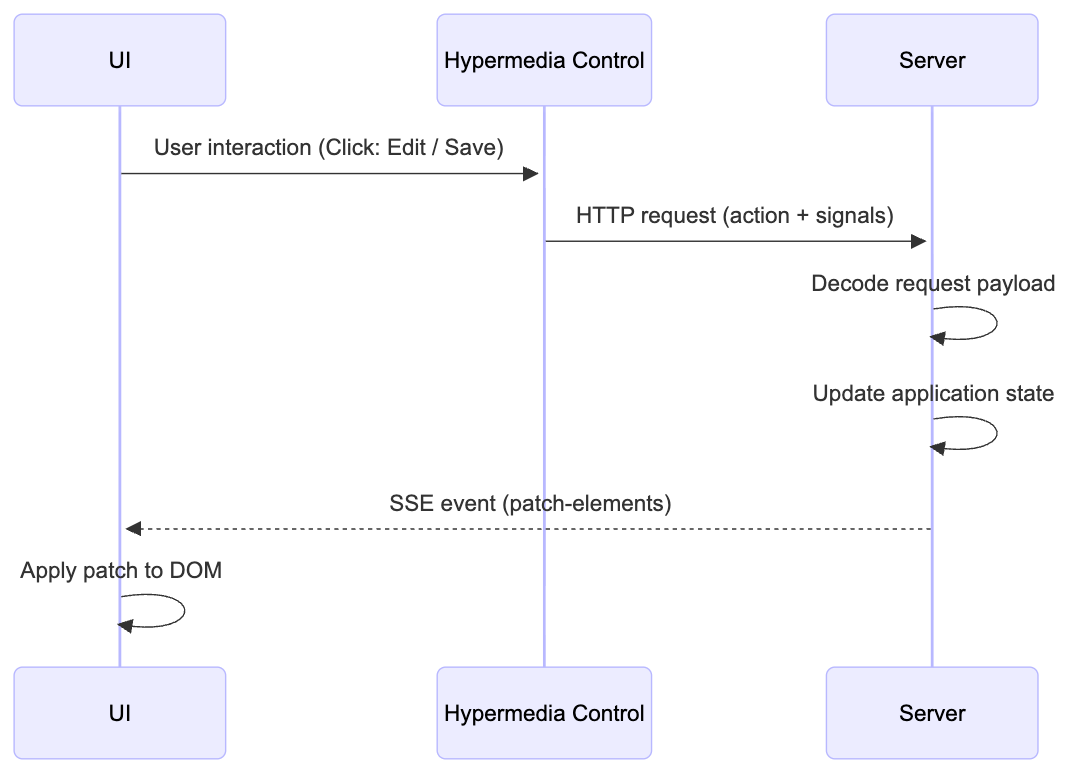}
  \caption{Signal-based formless submission interaction flow.}
  \label{click-to-edit-seq-diagram}
\end{figure}

Figure~\ref{click-to-edit-seq-diagram} illustrates the interaction flow of a
signal-driven formless submission pattern. The UI is structured around two
reactive view modes, read-only and edit, whose visibility and behavior are
governed by the server through HTML pacthes rather than traditional form
submissions or full page reloads.


%% file: sections/sec04-formal-def.tex
\section{Formalization of BRHC Control Patterns}
\label{sec:formal-signals}

To precisely model the \textit{Backend-driven Reactive Hypermedia Control (BRHC)
Patterns} within the established distributed hypermedia system $S = (F, P, C,
\text{SER})$, we introduce mathematical extensions to the foundational model
proposed by Gross et al.~\cite{gross2024hypermedia}. These extensions explicitly
account for persistent asynchronous reactive pipelines, fine-grained client-side
state storage, and signal-sharing primitives across the network boundary. Here,
the core system $S$ is formalized by the following distinct components:
\begin{itemize}
    \item $F$ is the complete set of data and media formats supported by the
    system (e.g., HTML fragments, JSON, text/event-stream).
    \item $P$ is the set of network protocols used for transmission and message
    exchange within the system architecture (e.g., HTTP, WebSockets).
    \item $C$ represents a conforming hypermedia client application (typically a
    web browser running a specialized hypermedia runtime engine) responsible for
    managing state and evaluating user input events.
    \item $\text{SER}$ represents a conforming hypermedia-driven network server
    that responds to client interactions exclusively via media types defined
    within $F$.
\end{itemize}

Network exchanges occur as standard request-response cycles or streams between
client $C$ and server $\text{SER}$ over a chosen protocol $P$, encoding payloads
explicitly in a specific format $F \in F$. Next, we present the mathematical
extensions to the foundational model:

\begin{itemize}
    \item Let $\mathcal{G}$ be the set of observable \textit{Signals}, where an
    individual signal $g \in \mathcal{G}$ represents a reactive, primitive state
    variable synchronized across the system. Each signal encapsulates a current
    scalar value $v \in \mathbb{V}$.
    \item Let $\mathcal{S}_{\mathcal{G}} \subseteq \mathcal{G}$ be the
    \textit{Client Signal Store} managed natively by the hypermedia execution
    engine embedded within client $C$. Consequently, the total composite user
    experience state of the client $UX$ is extended to include this store:
    \begin{equation}
        UX = (D_H, R(D_H), \mathcal{S}_{\mathcal{G}})
    \end{equation}
    where $D_H$ is the underlying hypermedia document structure and $R(D_H)$
    represents the client-side rendering projection.
    \item Let $P_{\text{stream}} \in P$ denote a persistent, asynchronous
    network protocol transport channel (e.g., Server-Sent Events) that
    facilitates a continuous, unidirectional push pipeline from server to
    client: $\vec{HE}: \text{SER} \xrightarrow{P_{\text{stream}}} C$.
\end{itemize}

Table~\ref{tab:dsl-mapping} summarizes the correspondence between the algebraic
components and their concrete realization in the HtmlFlow Kotlin DSL.
\textit{Signal Initialization} instantiates a type-safe signal in the
template compilation context.
\textit{Stream Connection} binds a client $C$ to the
server's event stream pipeline $P_{\text{stream}}$.
\textit{Interaction Control} hooks user actions ($\text{EVT}_{\text{action}}$) to
compiler-verified remote function entry points on $\text{SER}$.
\textit{Media Response} delivers optimized value state deltas omitting raw structural
layout presentation markup.
\textit{Reactive Bound Projection} subscribes a DOM text rendering node directly to the
lifecycle updates of a specific signal.

\begin{table}
\centering
\caption{Mapping of Formal Algebraic Components to the Kotlin DSL Abstractions.}
\label{tab:dsl-mapping}
{%
    \renewcommand{\texttt}[1]{{\ttfamily\fontsize{9}{11}\selectfont #1}}
    \begin{tabular}{ll}
    \toprule
    \textbf{Formal Component} & \textbf{Kotlin DSL Implementation} \\
    \midrule
    Signal Initialization ($g \in \mathcal{G}$) & \texttt{val count = dataSignal("count", 0)} \\
    Stream Connection ($\vec{HE}$) & \texttt{dataInit \{ get(::getEvents) \}} \\
    Interaction Control ($\text{ELT}$) & \texttt{button \{ dataOn(Click) \{ post(...) \} \}} \\
    Media Response ($M_{\text{signals}}$) & \texttt{event: datastar-patch-signals} \\
    Reactive Bound Projection & \texttt{dataText \{ +count \}} \\
    \bottomrule
    \end{tabular}
}
\end{table}


\subsection{Primitive Operational Phases}
\label{sec:primitive-phases}

The five BRHC patterns are composed of combinations of five primitive
operational phases. By decoupling these phases, we formalize how different
patterns reuse identical server-client interaction mechanics.

\medskip
\noindent \textbf{Phase I: Stream Initialization and Registration.} \\
When a specialized hypermedia control element $\text{ELT}_{\text{init}} \in D_H$
is mounted into the document structure by client $C$, a lifecycle initialization
event $\text{EVT}_{\text{mount}} \in \text{EVT}_C$ automatically executes an
asynchronous exchange to open the persistent data pipeline:
\begin{equation}
    \vec{HE}(C, P_{\text{stream}}, \text{SER}) \rightarrow \text{Establishes Open Reactive Channel}
\end{equation}
This channel maps an authoritative data feed on $\text{SER}$ to the client-side
store $\mathcal{S}_{\mathcal{G}}$, registering and instantiating a set of target
reactive signals $g \in \mathcal{S}_{\mathcal{G}}$.

\medskip
\noindent \textbf{Phase II: Client-Initiated Hypermedia Action.} \\
When a user-driven interaction event $\text{EVT}_{\text{action}} \in
\text{EVT}_C$ occurs on an actionable hypermedia control element $\text{ELT} \in
D_H$, it dispatches a discrete hypermedia exchange payload over a standard
protocol $P$ containing a snapshot of local parameters or the current signal
store state $\mathcal{S}_{\mathcal{G}}$:
\begin{equation}
    HE(C, P, \text{SER}) \quad \text{with payload } \mathcal{M}_{\text{req}} \in F
\end{equation}
Upon receiving this exchange, $\text{SER}$ executes authoritative application
logic and state mutations.

\medskip
\noindent \textbf{Phase III: Server-Driven Signal Push.} \\
Following a state mutation or an asynchronous server-side trigger, $\text{SER}$
utilizes the active asynchronous channel $\vec{HE}$ established during Phase I
to push a specific hypermedia signal frame, $M_{\text{signals}} \in F$ (e.g., a
\texttt{datastar-patch-signals} media fragment), directly targeting the client
signal store:
\begin{equation}
    \text{SER} \xrightarrow{\vec{HE}} M_{\text{signals}} \rightarrow C
\end{equation}

\medskip
\noindent \textbf{Phase IV: Signal Patch Disposition.} \\
Upon ingestion of $M_{\text{signals}}$, the hypermedia update function
$UP(\cdot)$ within $C$ executes under a specialized signal patch disposition,
denoted as $DI_{\text{patch}}$. This disposition updates the values inside the
client signal store $\mathcal{S}_{\mathcal{G}}$ while keeping the document $D_H$
structurally invariant:
\begin{equation}
    UP\left((D_H, R(D_H), \mathcal{S}_{\mathcal{G}}), \, M_{\text{signals}}, \, DI_{\text{patch}}\right) \rightarrow (D_H, R(D_H)', \mathcal{S}_{\mathcal{G}}')
\end{equation}
where $\mathcal{S}_{\mathcal{G}}' = \mathcal{S}_{\mathcal{G}} \setminus \{g
\mapsto v\} \cup \{g \mapsto v_{\text{new}}\}$. The modification of values
inside $\mathcal{S}_{\mathcal{G}}'$ causes the client-side reactive loop to
automatically re-evaluate presentation nodes bound to $g$, shifting the
rendering state $R(D_H) \to R(D_H)'$ without mutating DOM topology.

\medskip
\noindent \textbf{Phase V: Structural Fragment Transclusion.} \\
The server returns a standard structural hypermedia fragment payload
$M_{\text{fragment}} \in F$ containing raw HTML snippets accompanied by a
structural disposition directive $DI_{\text{merge}}$ (e.g., upsert, append, or
prepend). The client update function mutates the DOM topology directly:
\begin{equation}
    UP\left((D_H, R(D_H), \mathcal{S}_{\mathcal{G}}), \, M_{\text{fragment}}, \, DI_{\text{merge}}\right) \rightarrow (D_H', R(D_H'), \mathcal{S}_{\mathcal{G}})
\end{equation}
resulting in a structural layout mutation where $D_H \neq D_H'$.


\subsection{Composition Matrix of BRHC Patterns}
\label{sec:pattern-matrix}

Every BRHC pattern maps to a distinct configuration of the primitive operational
phases defined in Section \ref{sec:primitive-phases}. These patterns formalize
how the type-safe DSL orchestrates different combinations of state updates and
structural transclusions.

\medskip
\noindent \textbf{Pattern 1: Backend-Driven Signal Update}
This pattern is applied in the shared-counter Hypermedia Control presented in
Subsection~\ref{counter-pattern} and comprises the following primitive phases:
\begin{equation}
\text{Pattern}_1 = \text{Phase I} \rightarrow \text{Phase III} \rightarrow \text{Phase IV}
\end{equation}
\begin{itemize}
    \item Phase I: Stream Initialization and Registration.
    \item Phase III: Server-Driven Signal Push.
    \item Phase IV: Signal Patch Disposition.
\end{itemize}

\medskip
\noindent \textbf{Pattern 2: Action-Triggered HTML Patch}
This pattern handles layout updates that are triggered directly by a client
action. It is illustrated by the table-row deletion example presented in
Subsection~\ref{dsl-pattern2}, in which each row contains a delete button, and
comprises the following primitive phases:
\begin{equation}
\text{Pattern}_2 = \text{Phase II} \rightarrow \text{Phase V}
\end{equation}
\begin{itemize}
    \item Phase II: Client-Initiated Hypermedia Action.
    \item Phase V: Structural Fragment Transclusion.
\end{itemize}

\medskip
\noindent \textbf{Pattern 3: Signal-Based HTML Patch}
In this pattern, a client-side interaction triggers a backend operation that
uses the current signal state to generate an updated HTML fragment. The pattern
is illustrated by the active-search example of Subsection~\ref{dsl-pattern3},
where the user's search term is maintained as a signal and automatically
transmitted to the backend as the user types. The backend uses this value to
retrieve the matching contacts and responds with an HTML patch that updates the
results table. The pattern comprises the following primitive phases:

\begin{equation}
\text{Pattern}_3 = \text{Phase I} \rightarrow \text{Phase III} \rightarrow (\text{Phase IV} \parallel \text{Phase V})
\end{equation}
\begin{itemize}
    \item Phase I: Stream Initialization and Registration.
    \item Phase III: Server-Driven Signal Push.
    \item Phase IV: Signal Patch Disposition.
    \item Phase V: Structural Fragment Transclusion.
\end{itemize}

\medskip
\noindent \textbf{Pattern 4: Signal-Based Incremental HTML Patch}
This pattern establishes a continuous, asynchronous update cycle in which an
open server stream progressively delivers multiple HTML fragments over time,
allowing the DOM to be extended without requiring explicit user actions for each
update. It is illustrated in Subsection~\ref{dsl-pattern4} through an
infinite-scroll interaction model, where additional table rows are loaded and
appended as the user approaches the end of the currently visible content. The
pattern comprises the following primitive phases:

\begin{equation}
\text{Pattern}_4 = \text{Phase I} \rightarrow \text{Phase III} \rightarrow \text{Phase V}
\end{equation}
\begin{itemize}
    \item Phase I: Stream Initialization and Registration.
    \item Phase III: Server-Driven Signal Push.
    \item Phase V: Structural Fragment Transclusion.
\end{itemize}

\medskip
\noindent \textbf{Pattern 5: Signal-Based Formless Submission}

This pattern leverages the client signal store to transmit user inputs
asynchronously without traditional HTML form elements. A client action
serializes the current state signals, triggers an authoritative server mutation,
and receives updated state values via the open stream, which simultaneously 
updates client state variables and swaps targeted layout fragments.
The pattern comprises the following primitive phases:
\begin{equation}
\text{Pattern}_5 = \text{Phase II} \rightarrow \text{Phase III} \rightarrow \text{Phase V}
\end{equation}

\begin{itemize}
    \item Phase II: Client-Initiated Hypermedia Action.
    \item Phase III: Server-Driven Signal Push.
    \item Phase V: Structural Fragment Transclusion.
\end{itemize}

%% file: sections/sec05-case-study.tex
\section{PetClinic Case Study}
\label{sec:petclinic}

Here we analyze a case study based on the official Kotlin version of Spring
PetClinic. Spring PetClinic~\cite{petclinic} is an open-source application
originally developed in the Java ecosystem and commonly used to demonstrate web
development concepts, architectural patterns, and framework capabilities. It was
inspired by the PetStore application, which illustrated the use of J2EE
technologies for developing enterprise e-commerce web
applications~\cite{singh2002designing}. The application was selected as a case
study because it represents a realistic backend-driven web application with
conventional server-side rendering patterns, including controllers, domain models,
repositories, and HTML views. Its structure provides a representative baseline for
evaluating how existing web application workflows can be transformed into reactive
\textit{hypermedia controls} while preserving the backend as the primary source of
application state and interaction logic. The original application uses Thymeleaf
for server-side templates; in our case study, those views were reimplemented using
the HtmlFlow Datastar DSL, and the Spring MVC controllers were adapted to return
completed views, partial views and signal patches.

The migration changes how HTML representations are created and updated, with the
goal of demonstrating how a backend-driven application can express reactive
\textit{hypermedia controls} through a statically typed Kotlin DSL while following
a hypermedia-first approach. The first change concerns the owner search workflow.
In the original implementation, the user enters a last name in a form and submits
it as shown in Listing~\ref{lst-find-owners-kotlin-petclinic}. The browser then
issues a \texttt{GET} request to the \texttt{/owners} resource, with the last name
encoded as a query parameter. This interaction results in the server generating a
new HTML representation containing the query results, which replaces the current
document displayed by the browser.

\begin{lstlisting}[
  language=HTML,
  caption={Find Owners in Kotlin PetClinic},
  label={lst-find-owners-kotlin-petclinic},
  numbers=left
]
<form th:object="${owner}" th:action="@{/owners}" method="get">
    <div>
        <input th:field="*{lastName}"/>
    </div>
    <div>
        <button type="submit">Find Owner</button>
    </div>
</form>
\end{lstlisting}

Through the HtmlFlow Datastar API, the same workflow is redefined as an active
search, as shown in Listing~\ref{lst-find-owners-htmlflow-datastar}. In this
approach, the input element acts as a Hypermedia Control that triggers requests
during user interaction, while the backend response updates only the affected
results table fragment instead of replacing the complete page.

\begin{lstlisting}[
  language=kotlin,
  caption={Find Owners in HtmlFlow Datastar},
  label={lst-find-owners-htmlflow-datastar},
  numbers=left
]
val lastName = dataSignal(lastName)  
input {
    dataBind(lastName)
    dataOn(Input) {
        get("/owners/find")
    }
}
\end{lstlisting}

The input element in Listing~\ref{lst-find-owners-htmlflow-datastar} remains a
standard HTML element, but the DSL enriches it into a Hypermedia Control: the
\texttt{Input} event defines the interaction trigger,
\texttt{get("/owners/find")} specifies the HTTP request and target resource, and
the server response determines how the resulting fragment is integrated through
a Datastar patch operation.
The corresponding request handler decodes the Datastar signal payload included
with the request and extracts the current \texttt{lastName} value. This value is
then used to query the repository. Instead of generating and returning a
complete HTML page, the handler emits a Server-Sent Event containing an element
patch that updates only the table body fragment with the matching results.
The key change is not only that the interface becomes more interactive, but that
the interaction remains hypermedia-first. The server continues to define the
available operations, process the requests, and determine the resulting UI
representation, while the client is responsible only for applying the received
updates.

The same pattern appears in owner and pet edit operations, and in pet creation.
In the original application, these interactions navigated to separate pages,
requiring a full-page render for each. In the migrated version, they no longer
require navigation to separate pages. The controls send requests to the server
with the relevant signal states, and the server responds with partial element
patches that replace only the affected owner or pet region.
For example, the owner-detail page includes a control for initiating pet
creation, as shown in Listing~\ref{lst-add-pet-button-htmlflow-datastar}. The
corresponding request handler responds with an element patch that updates the
pets table by inserting a provisional row. This row contains signal-bound input
fields and a save control, as shown in
Listing~\ref{lst-save-pet-htmlflow-datastar}.

\begin{lstlisting}[
  language=kotlin,
  caption={Add Pet Button in HtmlFlow Datastar},
  label={lst-add-pet-button-htmlflow-datastar},
  numbers=left
]  
button {
    dataOn(Click) { get("/pets/new") }
    text("Add New Pet")
}
\end{lstlisting}

\begin{lstlisting}[
  language=kotlin,
  caption={Save Pet in HtmlFlow Datastar},
  label={lst-save-pet-htmlflow-datastar},
  numbers=left
]
val petName = dataSignal("petName")  
input { dataBind(petName) }
button {
    dataOn(Click) { post("/pets/new") }
    text("save")
}
\end{lstlisting}

When the save control is activated, the handler receives the current pet
signals (in this example \texttt{petName}) in the request body, validates and
persists the new pet, and returns another element patch. This patch replaces the
editing row with a default row for an existing pet.

These cases show how Reactive \textit{hypermedia controls} can provide interactive
behavior usually implemented with application-specific JavaScript, while
preserving Hypermedia as the Engine of Application State. The DSL reduces
complexity through statically-typed expressions for \textit{hypermedia controls},
following a server-driven approach.

%% file: sections/sec07-performance.tex
\section{Performance evaluation}
\label{sec:performance}

To evaluate the performance of the proposed approach, two complementary
benchmarking methodologies were employed. Lighthouse~\cite{lighthouse} was used to
measure initial page-load performance under a standardized browser environment,
while JMeter~\cite{jmeter}, together with the WebDriver Sampler plug-in and
Selenium WebDriver~\cite{selenium}, was used to benchmark complete interactive
workflows. Since both the Datastar and React implementations rely on client-side
JavaScript execution, browser-based measurements were required. Although
Playwright~\cite{playwright} supports browser automation, it does not provide
built-in metrics for throughput, network transfer, or repeated performance
measurements, making it less suitable for the comparative workload evaluation
conducted in this study.

The benchmark is based on selected use cases from the Spring PetClinic case
study presented in Section~\ref{sec:petclinic} and is publicly available in the
GitHub repository
\textit{spring-petclinic-benchmark}\footnote{\url{https://github.com/xmlet/spring-petclinic-benchmark}}.
The benchmark with JMeter covers two operations:  creating a pet, and finding owners of the
Spring PetClinic case study. The create pet operation exercises the Signal-based
Formless Submission pattern, and the find owners operation exercises the
Signal-based HTML patch pattern. These operations are measured across three
implementations (Thymeleaf~\cite{thymeleaf}, HtmlFlow Datastar, and
React~\cite{react}), each representing a different architectural paradigm.
Thymeleaf follows a traditional SSR model in which every user interaction triggers
a full server round-trip and a complete page reload. The HtmlFlow Datastar
implementation follows the Backend-Driven Reactive Hypermedia Control (BRHC)
patterns presented in Section~\ref{sec:patterns} by combining SSR with
transclusion-based updates. The initial response is fully rendered on the server
using HtmlFlow, while subsequent interactions are handled through
\textit{hypermedia controls} that trigger server-side processing and receive
partial HTML or signal updates via Datastar, which are transcluded into the
existing document without a full page reload. React implements a pure CSR model
following the SPA architecture, in which the server exposes a JSON API and all
view logic executes in the browser. The goal of this evaluation is to compare
their relative performance under identical conditions.

An automated script coordinates the full lifecycle: starting all server instances,
waiting for their availability, executing a warm-up phase of 25 iterations to
stabilise the runtime environment, and then running the main benchmark of 250
sequential iterations. The test plan is defined in a single JMeter 
\texttt{.jmx}\footnote{\url{https://github.com/xmlet/spring-petclinic-benchmark/tree/docker-setup/benchmark/jmeter}}
file covering the full workflow for all three implementations.
The benchmark intentionally uses a single execution thread, as introducing
concurrency would add variables that are unrelated to the comparison.
Multi-user concurrency is therefore deliberately excluded and reserved
for future work.

Each implementation runs inside a separate Docker~\cite{docker} container with
host networking and no resource limits, orchestrated through Docker Compose.
Service containers use OpenJDK (Eclipse Temurin)~\cite{openjdk} or
Node.js~\cite{nodejs}; the JMeter
container is based on Debian with Apache JMeter~\texttt{5.6.3} and Chromium.
All tests were conducted on a host machine with the specifications
listed in Table~\ref{tab:specs}.

\begin{table}
\centering
\caption{Host machine specifications.}
\label{tab:specs}
\begin{tabular}{@{}ll@{}}
\toprule
\textbf{Component} & \textbf{Specification} \\
\midrule
Host OS & Fedora Linux 44 (Workstation Edition) \\
CPU & 11th Gen Intel Core i5-1135G7 @ 2.40GHz \\
Cores / Threads & 4 cores / 8 threads \\
RAM & 15Gi \\
Disk & 150G NVMe \\
Kernel & 7.0.12-201.fc44.x86\_64 \\
\bottomrule
\end{tabular}
\end{table}

\subsection*{Initial load}
\label{sec:initial-load}

The initial-load benchmark was conducted using Lighthouse in navigation mode under
the desktop performance profile. The evaluated page corresponds to the owner
details view (\texttt{/owners/1}), which renders the owner's information, the
associated pets table, and the interactive controls for: 1) editing the owner, 2) adding
new pets, and 3) managing existing pets. This page was selected because it represents
one of the most feature-rich views in the application, combining server-rendered
content with multiple \textit{hypermedia controls}. The evaluation focuses on
three performance metrics: First Contentful Paint (FCP), which measures when the
first visible content is rendered; Largest Contentful Paint (LCP), which measures
when the main page content finishes rendering; and Speed Index (SI), which
estimates how quickly the page becomes visually complete.

The results are presented in Table~\ref{tab:results-initial}.

\begin{table}
\centering
\caption{Initial load results (Lighthouse on Google Chrome).}
\label{tab:results-initial}
\begin{tabular}{@{}lccc@{}}
\toprule
\textbf{Metric} & \textbf{Thymeleaf} & \textbf{HtmlFlow-Datastar} & \textbf{React} \\
\midrule
First Contentful Paint (ms)      & 606  & 610  & 1962  \\
Largest Contentful Paint (ms)    & 626  & 730  & 2056 \\
Speed Index (ms)                 & 606  & 610  & 1952 \\
\bottomrule
\end{tabular}
\end{table}

HtmlFlow-Datastar and Thymeleaf exhibit nearly identical initial-load
performance. Both deliver a fully server-rendered HTML document, with the only
additional client-side resource in the HtmlFlow-Datastar implementation being the
Datastar runtime (approximately 14\,kB). In contrast, the React implementation
must download and execute a JavaScript bundle of approximately 2\,MB before the
interface can be rendered.

Consequently, both HtmlFlow-Datastar and Thymeleaf achieve a First Contentful
Paint of 610\,ms, approximately $3.2\times$ faster than React (1962\,ms). Largest
Contentful Paint is similarly close, with Thymeleaf reaching 626\,ms and
HtmlFlow-Datastar 730\,ms, compared to 2056\,ms for React. The Speed Index follows
the same trend, with both server-rendered approaches reaching visual completeness
arround 600\,ms, whereas React requires 1952\,ms.

These results indicate that integrating Datastar into a server-rendered
application introduces negligible overhead during the initial page load. The
small additional runtime required by Datastar does not significantly affect page
rendering, allowing the HtmlFlow Datastar implementation to retain the initial
load characteristics of a traditional SSR application while providing the
reactive interaction model evaluated in the subsequent benchmarks.

\subsection*{Create Pet}

The create pet benchmark evaluates a representative user workflow, which
differs across implementations due to their respective architectural models:

\begin{itemize}
  \item \textbf{Thymeleaf (Multi-Page Application):} Each user action triggers a full page reload
  driven by the server.
  \begin{enumerate}
    \item Browser loads the owner details page as a full HTML document.
    \item User clicks ``Add New Pet'' $\rightarrow$ browser navigates to a new
    URL $\rightarrow$ server returns a full HTML form page.
    \item User fills the form and clicks ``Save'' $\rightarrow$ browser submits
    a form POST $\rightarrow$ server validates and persists the pet, returns a
    302 redirect.
    \item Browser follows the redirect $\rightarrow$ server returns the owner
    details page with the new pet.
  \end{enumerate}

  \item \textbf{HtmlFlow Datastar (Hypermedia-first):} The initial load
  returns a full HTML page. Subsequent interactions use Server-Sent Events to
  deliver partial DOM patches.
  \begin{enumerate}
    \item Browser loads the owner details page as a full HTML document containing
    Datastar client-side attributes.
    \item User clicks ``Add New Pet'' $\rightarrow$ a Datastar hypermedia
    control sends a GET request $\rightarrow$ server responds with an SSE
    stream containing an HTML fragment that is patched into the DOM, inserting
    the input row.
    \item User fills the input fields and clicks ``Save'' $\rightarrow$ the Datastar
    control sends a POST request with the form signals $\rightarrow$ server
    validates, persists the pet, and responds with an SSE stream that replaces
    the input row with the new pet row.
  \end{enumerate}

  \item \textbf{React (SPA):} The client loads the application entry point, then
  communicates with the server exclusively through JSON API calls. The view is
  rendered entirely on the client.
  \begin{enumerate}
    \item Browser loads the SPA entry point (HTML + JavaScript bundles) from the
    frontend server. React Router renders the owner details component, which
    fetches \texttt{/api/owners/\{id\}} and renders the view from the JSON
    response.
    \item User clicks ``Add New Pet'' $\rightarrow$ React Router handles the
    navigation client-side $\rightarrow$ the new component fetches
    \texttt{/api/pettypes} and \texttt{/api/owners/\{id\}} in parallel, then
    renders the form.
    \item User fills the form and clicks ``Save'' $\rightarrow$ client sends a
    JSON POST to the server $\rightarrow$ server validates and persists the pet,
    returns 201 $\rightarrow$ client navigates back to the owner details and
    re-fetches the owner data to display the new pet.
  \end{enumerate}
\end{itemize}

The JMeter measurement results are presented in Table~\ref{tab:results-create-pet}.

\begin{table}[h]
\centering
\caption{Create pet results (1 thread, 250 runs, Selenium WebDriver with Chromium).}
\label{tab:results-create-pet}
\begin{tabular}{@{}lrrr@{}}
\toprule
\textbf{Metric} & \textbf{Thymeleaf} & \textbf{HtmlFlow} & \textbf{React} \\
\midrule
Avg (ms)     & 732.23 & 571.22 & 1220.66 \\
Std Dev (ms) & 230.89 & 263.17 & 263.96 \\
Network (KB/req) & 11.98 & 19.57 & 186.38 \\
\bottomrule
\end{tabular}
\end{table}

HtmlFlow Datastar achieved the lowest latency, averaging 571\,ms, making it
1.28$\times$ faster than Thymeleaf (732\,ms) and 2.14$\times$ faster than React
(1221\,ms). This result is particularly noteworthy given that the three
implementations employ fundamentally different interaction models. Thymeleaf
performs three full page loads throughout the workflow (owner details, pet form,
and owner details after redirect). In contrast, HtmlFlow Datastar performs a
single full page load followed by two SSE exchanges that patch the DOM in place.
React, on the other hand, relies on multiple sequential API interactions combined
with client-side rendering.
HtmlFlow Datastar exhibited similar variability to the other implementations,
with a standard deviation of 263\,ms, compared to 231\,ms for Thymeleaf and
264\,ms for React. 

The network transfer volume differed across implementations.
HtmlFlow Datastar averaged 19.57\,KB per request, compared to
11.98\,KB for Thymeleaf and 186.38\,KB for React.
In all three implementations, the number of pets displayed in the owner details
view was limited to the 10 most recent, preventing unbounded page growth as
pets accumulated during the benchmark.

\subsection*{Find Owners}

The third benchmark assesses the find owners operation, where there are also clear
differences in implementation and workflow:

\begin{itemize}
  \item \textbf{Thymeleaf (Multi-Page Application):} Each user action triggers a full page reload
  driven by the server.
  \begin{enumerate}
    \item Browser loads the find owners page as a full HTML document.
    \item User enters a search string to filter by owner last name.
    \item User clicks ``Find Owner'' button $\rightarrow$ browser triggers a GET
    request $\rightarrow$ server returns the new page with filtered owners.
  \end{enumerate}

  \item \textbf{HtmlFlow Datastar (Hypermedia-first):} The initial load
  returns a full HTML page. Subsequent interactions use Server-Sent Events to
  deliver partial DOM patches.
  \begin{enumerate}
    \item Browser loads the find owners page as a full HTML document containing
    Datastar client-side attributes. The owner table is initially empty.
    \item User enters a search string to filter by owner last name $\rightarrow$
    Datastar hypermedia control automatically sends a GET request $\rightarrow$
    server responds with an SSE stream containing an HTML fragment that is
    patched into the DOM, replacing the empty table body with the matching
    results.
  \end{enumerate}

  \item \textbf{React (SPA):} The client loads the application entry point, then
  communicates with the server exclusively through JSON API calls. The view is
  rendered entirely on the client.
  \begin{enumerate}
    \item Browser loads the SPA entry point (HTML + JavaScript bundles) from the
    frontend server. React Router renders the find owners component, which
    fetches \texttt{/api/owners} and renders the full owner list from the JSON
    response.
    \item User enters a search string to filter by owner last name.
    \item User clicks ``Find Owner'' button $\rightarrow$ the client sends a
    GET request to the JSON API $\rightarrow$ the server returns a JSON array
    with the matching owners $\rightarrow$ React re-renders the table with the
    filtered results.
  \end{enumerate}
\end{itemize}

The JMeter measurement results are presented in Table~\ref{tab:results-find-owners}.

\begin{table}[h]
\centering
\caption{Find owners results (1 thread, 250 runs, Selenium WebDriver with Chromium).}
\label{tab:results-find-owners}
\begin{tabular}{@{}lrrr@{}}
\toprule
\textbf{Metric} & \textbf{Thymeleaf} & \textbf{HtmlFlow} & \textbf{React} \\
\midrule
Avg (ms)     & 595.46 & 166.07 & 695.08 \\
Std Dev (ms) & 223.63 & 184.26 & 19.07 \\
Network (KB/req) & 3.85 & 3.80 & 181.55 \\
\bottomrule
\end{tabular}
\end{table}

HtmlFlow Datastar achieved the lowest latency, averaging
166\,ms, 3.6$\times$ faster than Thymeleaf (595\,ms) and 4.2$\times$ faster than
React (695\,ms). This advantage reflects the efficiency of the SSE-based
hypermedia model: the input event triggers a single lightweight SSE request that
patches only the table body, whereas Thymeleaf requires a complete page reload
and React incurs the overhead of client-side rendering and JSON API
serialisation. React exhibited the lowest variability, with a
standard deviation of 19\,ms, compared to 184\,ms for HtmlFlow Datastar and
224\,ms for Thymeleaf. This reflects the SPA model's consistent client-side
rendering cost for a simple table update, independent of server-side processing
time.
Regarding network transfer, Thymeleaf transferred 3.85\,KB and HtmlFlow Datastar
transferred 3.80\,KB per request (server-rendered HTML tables), while React's
network volume was 181.55\,KB per request.

\subsection*{Summary}

The results indicate that HtmlFlow Datastar consistently achieved the best overall
performance across the evaluated interaction benchmarks, delivering the lowest
latency while sustaining the highest throughput among the three implementations.
Regarding the initial page load, Lighthouse showed that HtmlFlow Datastar
exhibited performance comparable to Thymeleaf, with nearly identical First
Contentful Paint, Largest Contentful Paint, and Speed Index values. This indicates
that integrating the Datastar runtime introduces negligible overhead during the
initial rendering of a server-rendered page. React, in contrast, exhibited
substantially higher rendering times, reflecting the additional cost of
downloading and executing the client-side application bundle before the interface
becomes interactive. During the interaction benchmarks, React exhibited
the highest latency and lowest throughput in the create pet benchmark. In the find
owners benchmark, React again exhibited the highest latency, followed by Thymeleaf,
with HtmlFlow Datastar achieving the lowest. In terms of network transfer, Thymeleaf and HtmlFlow
Datastar transferred modest volumes in both workflows (3.85\,KB and 3.80\,KB per request in
find owners; 11.98\,KB and 19.57\,KB per request in create pet), reflecting the efficiency
of server-rendered HTML. React's network volume was substantially higher across
both benchmarks (181.55\,KB and 186.38\,KB per request).

The benchmark was intentionally executed using a single sequential thread to
isolate per-request latency and throughput. Concurrency behavior depends heavily
on the underlying server execution model, whether based on thread pools or
event-loop architectures, and these characteristics vary across the evaluated
implementations. Including such factors in the analysis would conflate
application-level interaction costs with server-side scheduling behavior.
Consequently, the scope of this evaluation is limited to single-user interaction
performance and does not address multi-user workloads, resource contention, or
scalability under concurrent access, which would require a separate study.

Despite these limitations, the results support the central hypothesis of this
work: a backend-driven, hypermedia-first architecture based on reactive signals
can achieve competitive performance while avoiding some of the latency and
throughput overheads commonly associated with SPA architectures. At the same time,
it preserves the simplicity of a server-centric programming model, where
application state, interaction logic, and UI updates remain primarily coordinated
by the backend.

%% file: sections/sec08-discussion.tex
\section{Discussion}
\label{sec:disc}

\subsection{Interpretation of the results}

The results indicate that Backend-driven Reactive \textit{hypermedia controls} can
be effectively expressed through a statically typed Kotlin DSL without relying on
application-level JavaScript expressions. More importantly, the approach shifts a
substantial part of the interaction definition from stringly typed HTML
attributes into compiler-checked Kotlin constructs. Existing libraries such as
htmx and Datastar make server-driven interaction practical, but their native
programming model still depends on string-based signal names, URL references,
selectors, event declarations, and inline expressions. These mechanisms are
flexible, but difficult to verify before execution. The primary benefit of the
proposed DSL is therefore not only the reduction of handwritten JavaScript, but
also the movement of reactive hypermedia control definitions into a statically
typed environment where signal bindings, backend handler references, and
expression composition can be checked, refactored, and reused.

The practical applicability of these ideas was evaluated through the Spring
PetClinic migration and the performance benchmark. The migration demonstrated
that workflows originally implemented through Thymeleaf templates and full-page
navigation can be reformulated as reactive \textit{hypermedia controls} based on
partial updates. Furthermore, the benchmark showed lower latency and higher
throughput than the evaluated Thymeleaf and React implementations. Together,
these findings suggest that backend-driven reactive architectures can provide
both expressive interaction models and competitive runtime performance.

The main trade-off is that the DSL shifts complexity from handwritten
client-side scripts and external templates into the backend type system and DSL
design. This trade-off is most relevant in contexts where compiler support,
backend ownership of state, and reduced frontend/backend heterogeneity are
desirable architectural properties. It may be less suitable when teams require
the flexibility of dedicated frontend frameworks or when significant portions of
the interaction logic are naturally client-local.

The approach also has practical limitations. Since it is fundamentally
server-driven and relies on HTTP requests and, in several patterns, Server-Sent
Events, it assumes network connectivity for authoritative state transitions.
Although Service Workers~\cite{mdn-service-worker} could provide partial offline
support through asset caching and deferred synchronization, workflows involving
server-pushed updates, conflict resolution, or strongly consistent state remain
challenging. Offline capabilities should therefore be regarded as future work
rather than an inherent property of the proposed model.

\subsection{Threats to Validity}

The evaluation presented in this work may be subject to threats to validity.
Although the pattern catalog, the Spring PetClinic migration, and the performance
evaluation provide evidence supporting the proposed approach, the results should
be interpreted within the scope of the adopted methodology, implementation, and
experimental setting.

\textbf{Construct validity.} The evaluation focuses on two research questions:
the elimination of application-level JavaScript expressions and the improvement
of type safety through a statically typed DSL. While the pattern catalog and
Petclinic migration provide evidence that these goals can be achieved, the study
does not directly measure developer productivity, maintainability, defect rates,
or comprehension. Consequently, improvements in these dimensions are inferred
rather than empirically demonstrated.

\textbf{Internal validity.} The implementation and evaluation are based on a
specific technology stack consisting of Kotlin, HtmlFlow, Spring MVC, and
Datastar. The observed results may therefore be influenced by characteristics of
these technologies rather than exclusively by the DSL design itself. In
addition, the performance benchmarks were conducted in a controlled environment
and primarily reflect single-user workloads, limiting the ability to attribute
all observed performance differences solely to the architectural model.

\textbf{External validity.} The pattern catalog covers representative reactive
\textit{hypermedia controls} but is not exhaustive. Furthermore, the Petclinic
migration represents a single case study in a well-understood domain. The results
may not generalize to larger systems, domains with substantially different
interaction requirements, or organizations with different frontend/backend
responsibilities.

%% file: sections/sec09-conclusions.tex
\section{Conclusions}
\label{sec:conclusions}

This paper presented an approach for expressing Backend-driven Reactive
\textit{hypermedia controls} through a statically typed Kotlin DSL built on top of
HtmlFlow and Datastar. The central idea is to move Hypermedia Control
definitions, signal bindings, and request composition from string-based HTML
attributes into compiler-checked Kotlin constructs, while preserving a
hypermedia-first interaction model in which the server remains the authoritative
source of application state and UI evolution. To support this model, we defined a
catalog of Backend-driven Reactive Hypermedia Control patterns, showed how they
can be expressed without application-specific JavaScript expressions, and
formalized their structure in terms of signals, triggers, requests, and server
emitted updates.

The first research question asked to what extent \textit{hypermedia controls} and
reactive signal bindings can be expressed through a statically typed Kotlin DSL
while eliminating application-level JavaScript expressions. The results show
that the proposed DSL is expressive enough to capture the studied reactive
interaction patterns without requiring application-specific JavaScript in the
view layer. The runtime still depends on Datastar in the browser to interpret
generated attributes, maintain client-side signal state, and apply
Server-Sent Event patches, but the interaction logic itself is authored in
Kotlin rather than in stringly typed client-side expressions.

The second research question examined whether this approach improves type safety
and unifies frontend and backend programming models. The proposed DSL contributes
to both objectives by allowing signal bindings, backend handler references, and
HTML construction to coexist within the same statically typed language and
toolchain. Furthermore, the Spring PetClinic case study demonstrated that
workflows originally implemented with Thymeleaf templates and full-page
navigation can be migrated to a backend-driven reactive model based on partial
HTML updates and signal-aware \textit{hypermedia controls}.

Taken together, these results indicate that hypermedia-first architectures
remain a viable alternative to SPA-centric development for a substantial class
of interactive web applications. By combining server-driven interactions with a
statically typed DSL, the proposed approach reduces frontend/backend
heterogeneity while preserving competitive runtime performance and a
compiler-assisted development experience.

Several future directions follow naturally from this work. First, the approach
should be evaluated on larger and more heterogeneous applications to assess its
scalability across richer interaction models and development teams. Second,
stronger verification mechanisms could be explored for selectors, patch targets,
and signal payload contracts. Third, comparative studies with template-based and
SPA-based implementations are needed to directly evaluate developer productivity,
maintainability, and comprehension. Fourth, future work could investigate
offline-capable variants based on Service Workers, deferred synchronization, and
conflict-resolution strategies. Finally, an important direction is the
integration of this DSL into AI-assisted web development workflows, where
statically typed hypermedia control definitions could serve as a structured and
verifiable target for code generation, reasoning, and automatic refactoring in
LLM-based development tools, potentially improving both correctness and
consistency of generated web interactions.

%% file: bibliography.bib
@article{carvalho2020,
	author="Carvalho, Fernando Miguel and Duarte, Luis and Gouesse, Julien",
	editor="Bozzon, Alessandro and Dom{\'i}nguez Mayo, Francisco Jos{\'e} and Filipe, Joaquim",
	title="Text Web Templates Considered Harmful",
	journal="Lecture Notes in Business Information Processing",
	year="2020",
	publisher="Springer International Publishing",
	address="Cham",
	pages="69--95",
	isbn="978-3-030-61750-9"
}

@book{Fowler02,
 author = {Fowler, Martin},
 title = {Patterns of Enterprise Application Architecture},
 year = {2002},
 isbn = {0321127420},
 publisher = {Addison-Wesley Longman Publishing Co., Inc.},
 address = {Boston, MA, USA},
}

@techreport{pebble,
  author = {Pebble},
  title = {Pebble},
  year = {2013},
  institution = {Pebble Templates},
  note   = {Accessed 2026-07-09},
  address={\url{https://github.com/PebbleTemplates/pebble}}
}

@techreport{handlebars,
  author = {Yehuda Katz},
  title = {Handlebars},
  year={2023},
  institution = {Handlebars.js},
  address={\url{http://handlebarsjs.com/}},
  note   = {Accessed 2026-07-09}
}

@techreport{thymeleaf,
	author = {Daniel Fernández},
	title = {Thymeleaf},
	year = {2011},
  institution = {Thymeleaf Project},
	address={\url{https://www.thymeleaf.org/}},
  note   = {Accessed 2026-07-09},
}

@techreport{react,
	author = {Jordan Walke},
	title = {React JavaScript library for building user interfaces},
	year = {2013},
  institution = {Meta},
	address={\url{https://reactjs.org/}},
  note   = {Accessed 2026-07-09},
}

@techreport{jmeter,
  title={Using jmeter to performance test web services},
  author={Nevedrov, Dmitri},
  journal={Published on dev2dev},
  pages={1--11},
  year={2006},
  institution={Oracle},
  publisher={Citeseer},
  address={\url{https://www.oracle.com/technical-resources/articles/enterprise-architecture/jmeter-performance-testing-part1.html}},
  note   = {Accessed 2026-07-09}
}

@article{nelson1995heart,
  title={The heart of connection: hypermedia unified by transclusion},
  author={Nelson, Theodor Holm},
  journal={Communications of the ACM},
  volume={38},
  number={8},
  pages={31--33},
  year={1995},
  publisher={ACM New York, NY, USA}
}

@article{krottmaier2001,
  title={Transclusions in the 21st Century.},
  author={Krottmaier, Harald and Maurer, Hermann A},
  journal={J. Univers. Comput. Sci.},
  volume={7},
  number={12},
  pages={1125--1136},
  year={2001}
}

@article{maurer2006,
  title={Transclusions in an html-based environment},
  author={Maurer, Hermann and Kolbitsch, Josef},
  journal={Journal of Computing and Information Technology},
  volume={14},
  number={2},
  pages={161--173},
  year={2006},
  publisher={Sveu{\v{c}}ili{\v{s}}te u Zagrebu Sveu{\v{c}}ili{\v{s}}ni ra{\v{c}}unski centar}
}

@book{fielding2000architectural,
  title={Architectural styles and the design of network-based software architectures},
  author={Fielding, Roy Thomas},
  year={2000},
  publisher={University of California, Irvine}
}

@inbook{vepsalainen2023desapearingframeworks,
  author = {Vepsäläinen, Juho and Hellas, Arto and Vuorimaa, Petri},
  year = {2023},
  month = {06},
  pages = {319-326},
  title = {The Rise of Disappearing Frameworks in Web Development},
  isbn = {978-3-031-34443-5},
  publisher = {Springer},
  doi = {10.1007/978-3-031-34444-2_23}
}

@article{vepsalainen2026revisiting,
  title={Revisiting Hypermedia, The Forgotten Web Application Development Paradigm},
  author={Veps{\"a}l{\"a}inen, Juho},
  year={2026},
  journal={TechRxiv},
  publisher={TechRxiv}
}

@inproceedings{gross2024hypermedia,
  title={Hypermedia controls: Feral to formal},
  author={Gross, Carson and Shaffer, Dillon and Revelle, Matt},
  booktitle={Proceedings of the 35th ACM Conference on Hypertext and Social Media},
  pages={52--64},
  year={2024},
  organization={ACM}
}

@book{gross2023hypermedia,
  author    = {Gross, Carson and Stepinski, Adam and Ak\c{s}im\c{s}ek, Deniz},
  title     = {Hypermedia Systems},
  year      = {2023},
  publisher = {Self-published},
  url       = {https://hypermedia.systems},
  note      = {Available free online. Accessed: 2026-04-07}
}

@article{sireteanu2021front,
  title={Front-end Frameworks for Development of SPA and MPA Web Applications},
  author={Sireteanu, Napoleon-Alexandru and Homocianu, Daniel},
  journal={Available at SSRN 3987838},
  year={2021}
}

@misc{ITUFactsFigures2025,
  author = {{International Telecommunication Union}},
  title = {Facts and Figures 2025},
  year = {2025},
  howpublished = {Online report},
  url = {https://www.itu.int/itu-d/reports/statistics/facts-figures-2025/}
}

@misc{W3TechsJavaScript,
  author = {{W3Techs}},
  title = {Usage Statistics of Client-Side Programming Languages for Websites},
  year = {2026},
  howpublished = {Online statistics},
  url = {https://w3techs.com/technologies/overview/client_side_language}
}

@misc{HTTPArchive2025,
  author = {{HTTP Archive}},
  title = {The 2025 Web Almanac},
  year = {2025},
  howpublished = {Online report},
  url = {https://almanac.httparchive.org/en/2025/}
}

@misc{Miller2020,
  author       = {Jason Miller},
  title        = {Islands Architecture},
  year         = {2020},
  howpublished = {\url{https://jasonformat.com/islands-architecture/}},
}

@misc{HallieOsmani2022,
  author       = {Hallie, L. and Osmani, A.},
  title        = {Islands Architecture},
  year         = {2022},
  howpublished = {\url{https://www.patterns.dev/posts/islands-architecture/}},
}

@inproceedings{mikkonen2008web,
  title={Web applications--spaghetti code for the 21st century},
  author={Mikkonen, Tommi and Taivalsaari, Antero},
  booktitle={2008 Sixth international conference on software engineering research, management and applications},
  pages={319--328},
  year={2008},
  organization={IEEE}
}

@techreport{datastar,
  author = {StarFederation},
  title = {Datastar: The Hypermedia Framework},
  year = {2024},
  institution = {Star Federation},
  address = {\url{https://data-star.dev/}},
  note   = {Accessed 2026-07-09}
}

@techreport{htmx,
  author = {Gross, Carson},
  title = {htmx: High Power Tools for {HTML}},
  year = {2020},
  institution = {Big Sky Software},
  address = {\url{https://htmx.org}},
  note   = {Accessed 2026-07-09},
}

@techreport{turbo,
  author = {Basecamp},
  title = {Hotwire: The Fast Way to Build Web Applications},
  year = {2020},
  institution = {Basecamp},
  address = {\url{https://hotwire.dev}},
  note   = {Accessed 2026-07-09}
}

@techreport{htmlflow,
  author= {Gamboa, Miguel},
  title= {HtmlFlow: Java DSL for Typesafe HTML},
  year= {2017},
  institution = {HtmlFlow},
  address = {\url{https://htmlflow.org}},
  note   = {Accessed 2026-07-09},
}

@techreport{nextjs,
  author = {{Vercel}},
  title = {Next.js},
  year = {2016},
  institution = {Vercel},
  address = {\url{https://nextjs.org}},
  note   = {Accessed 2026-07-09},
}

@book{geers2020micro,
  title={Micro frontends in action},
  author={Geers, Michael},
  year={2020},
  publisher={Simon and Schuster}
}

@misc{nextjs_isr_guide,
  author       = {{Vercel}},
  title        = {How to implement Incremental Static Regeneration (ISR)},
  year         = {2026},
  howpublished = {Documentation page},
  address = {\url{https://nextjs.org/docs/app/guides/incremental-static-regeneration}},
  note         = {Last Update: 2026-03-25. Accessed: 2026-06-05}
}

@techreport{ kotlinlang,
	title={Kotlin Language Documentation},
	author={Andrey Breslav},
	year={2024},
  institution={JetBrains},
	address={\url{https://kotlinlang.org/docs/kotlin-docs.pdf}},
	note   = {Accessed 2026-07-09},
}

@techreport{mdn-sse,
  author = {{MDN Web Docs}},
  title = {Server-sent events},
  year = {2015},
  institution = {MDN},
  address = {\url{https://developer.mozilla.org/en-US/docs/Web/API/Server-sent_events}},
  note   = {Accessed 2026-07-09},
}

@techreport{state-flow-kotlin,
  author = {JetBrains},
  title = {Kotlin StateFlow Documentation},
  year = {2024},
  institution = {JetBrains},
  address = {\url{https://kotlinlang.org/api/kotlinx.coroutines/kotlinx-coroutines-core/kotlinx.coroutines.flow/-state-flow/}},
  note   = {Accessed 2026-07-09}
}

@article{elliott1997functional,
  title={Functional Reactive Animation},
  author={Elliott, Conal and Hudak, Paul},
  journal={International Conference on Functional Programming (ICFP)},
  year={1997}
}

@article{frp_bainomugisha2013survey,
  title={A Survey on Reactive Programming},
  author={Bainomugisha, Engineer and others},
  journal={ACM Computing Surveys},
  volume={45},
  number={4},
  year={2013}
}

@article{carpenter2021solidjs,
  title={SolidJS: Reactive UI Library with Fine-Grained Reactivity},
  author={Carpenter, Ryan},
  journal={Technical Report / Project Documentation},
  year={2021}
}

@misc{tc39signals,
  author       = {{ECMAScript Technical Committee (TC39)}},
  title        = {Signals Proposal},
  year         = {2025},
  howpublished = {\url{https://github.com/tc39/proposal-signals}},
  note         = {Stage 1 Proposal, accessed 2026-06-11}
}

@book{kleppmann2017designing,
  title     = {Designing Data-Intensive Applications},
  author    = {Kleppmann, Martin},
  year      = {2017},
  publisher = {O'Reilly Media},
  isbn      = {978-1449373320},
  chapter   = {Time and Stream Processing},
  note      = {Discusses stream processing and continuously updated state in reactive systems}
}

@inproceedings{taivalsaari2017web,
  title={The web as a software platform: Ten years later},
  author={Taivalsaari, Antero and Mikkonen, Tommi},
  booktitle={International conference on web information systems and technologies},
  volume={2},
  pages={41--50},
  year={2017},
  organization={SCITEPRESS}
}

@misc{jquery,
  title        = {jQuery},
  author       = {{OpenJS Foundation}},
  year         = {2026},
  howpublished = {\url{https://jquery.com}},
  note         = {Accessed: 2026-06-12}
}

@article{mesbah2008,
author = {Mesbah, Ali and van Deursen, Arie},
title = {A component- and push-based architectural style for ajax applications},
year = {2008},
issue_date = {December, 2008},
publisher = {Elsevier Science Inc.},
address = {USA},
volume = {81},
number = {12},
issn = {0164-1212},
url = {https://doi.org/10.1016/j.jss.2008.04.005},
doi = {10.1016/j.jss.2008.04.005},
journal = {J. Syst. Softw.},
month = dec,
pages = {2194–2209},
numpages = {16}
}

@book{singh2002designing,
  title={Designing enterprise applications with the J2EE platform},
  author={Singh, Inderjeet},
  year={2002},
  publisher={Addison-Wesley Professional}
}

@misc{petclinic,
  author       = {{Spring Team}},
  title        = {Spring PetClinic Sample Application},
  year         = {2025},
  howpublished = {\url{https://spring-petclinic.github.io}},
  note         = {Open-source reference application demonstrating Spring-based web application development}
}

@techreport{mdn-service-worker,
  author = {{MDN Web Docs}},
  title = {Service Worker},
  year = {2015},
  institution = {MDN},
  address = {\url{https://developer.mozilla.org/en-US/docs/Web/API/Service_Worker_API}},
  note   = {Accessed 2026-07-09},
}

@misc{htmxdocs,
  author       = {{Big Sky Software}},
  title        = {htmx ~ Documentation},
  howpublished = {\url{https://htmx.org/docs/}},
  year         = {2026},
  note         = {Accessed: 2026-06-17}

}

@misc{datastardocs,
  author       = {{Star Federation}},
  title        = {Datastar ~ Guide},
  howpublished = {\url{https://data-star.dev/guide/getting_started}},
  year         = {2026},
  note         = {Accessed: 2026-06-17}

}

@techreport{selenium,
  author = {{Software Freedom Conservancy}},
  title = {Selenium WebDriver},
  year = {2004},
  institution = {Software Freedom Conservancy},
  address = {\url{https://www.selenium.dev/}},
  note   = {Accessed 2026-07-09},
}

@techreport{lighthouse,
  author = {{Google}},
  title = {Lighthouse: Automated Auditing, Performance Metrics, and Best Practices for the Web},
  year = {2016},
  institution = {Google},
  address = {\url{https://developer.chrome.com/docs/lighthouse/}},
  note   = {Accessed 2026-07-09},
}

@techreport{playwright,
  author = {{Microsoft}},
  title = {Playwright: Fast and Reliable End-to-End Testing for Modern Web Apps},
  year = {2020},
  institution = {Microsoft},
  address = {\url{https://playwright.dev/}},
  note   = {Accessed 2026-07-09},
}

@techreport{openjdk,
  author = {{Eclipse Foundation}},
  title = {Eclipse Temurin: Open Source Java SE Builds},
  year = {2021},
  institution = {Eclipse Foundation},
  address = {\url{https://adoptium.net/}},
  note   = {Accessed 2026-07-09},
}

@techreport{nodejs,
  author = {{OpenJS Foundation}},
  title = {Node.js},
  year = {2009},
  institution = {OpenJS Foundation},
  address = {\url{https://nodejs.org/}},
  note   = {Accessed 2026-07-09},
}

@techreport{docker,
  author = {{Docker Inc.}},
  title = {Docker: Accelerated Container Application Development},
  year = {2013},
  institution = {Docker Inc.},
  address = {\url{https://www.docker.com/}},
  note   = {Accessed 2026-07-09},
}
